\documentclass[namedreferences]{SolarPhysics}
\usepackage[optionalrh]{spr-sola-addons} % For Solar Physics
\usepackage{graphicx}
\usepackage{color}
\usepackage{url}             % For breaking URLs easily trough lines
\newcommand{\adv}{    {\it Adv. Spa. Res.}}

\newcommand{\aap}{    {\it Astron. Astrophys.}}

\newcommand{\ag}{     {\it Ann. Geophys.}}

\newcommand{\apj}{    {\it Astrophys. J.}}

\newcommand{\grl}{    {\it Geophys. Res. Lett.}}

\newcommand{\jgr}{    {\it J. Geophys. Res.}}
\newcommand{\mnras}{  {\it Mon. Not. Roy. Astron. Soc.}}

\newcommand{\pasj}{   {\it Pub. Astron. Soc. Japan}}

\newcommand{\solphys}{{\it Solar Phys.}}
\newcommand{\sovast}{ {\it Sov. Astronom.}}
\newcommand{\ssr}{    {\it Space Sci. Rev.}}

\begin{document}
\begin{article}
\begin{opening}

\title{A Challenging Solar Eruptive Event of 18 November 2003
and the Causes of the 20 November Geomagnetic Superstorm.
I.~Unusual History of an Eruptive Filament}

\author{V.V.~\surname{Grechnev}$^{1}$\sep
        A.M.~\surname{Uralov}$^{1}$\sep
        V.A.~\surname{Slemzin}$^{2}$\sep
        I.M.~\surname{Chertok}$^{3}$\sep
        B.P.~\surname{Filippov}$^{3}$\sep
        G.V.~\surname{Rudenko}$^{1}$\sep
        M.~\surname{Temmer}$^{4}$}

\runningauthor{Grechnev et al.} \runningtitle{Unusual History of
an Eruptive Filament on 18 November 2003}

\institute{$^{1}$ Institute of Solar-Terrestrial Physics SB RAS,
Lermontov St.\ 126A, Irkutsk 664033, Russia
                  email: \url{grechnev@iszf.irk.ru} email: \url{uralov@iszf.irk.ru}\\
           $^{2}$ P.N. Lebedev Physical Institute, Leninsky Pr., 53, Moscow, 119991,
                  Russia
                  email: \url{slem@lebedev.ru}\\
           $^{3}$ Pushkov Institute of Terrestrial Magnetism,
                  Ionosphere and Radio Wave Propagation (IZMIRAN), Troitsk, Moscow
                  Region, 142190 Russia
                  email: \url{ichertok@izmiran.ru}\\
           $^{4}$ IGAM/Kanzelh{\"o}he Observatory, Institute of Physics, UNI Graz
                  email: \url{manuela.temmer@uni-graz.at}}

\date{Received ; accepted }

\begin{abstract}

This is the first of four companion papers, which comprehensively
analyze a complex eruptive event of 18 November 2003 in active
region (AR) 10501 and the causes of the largest Solar Cycle 23
geomagnetic storm on 20 November 2003. Analysis of a complete data
set, not considered before, reveals a chain of eruptions to which
hard X-ray and microwave bursts responded. A filament in AR 10501
was not a passive part of a larger flux rope, as usually
considered. The filament erupted and gave origin to a coronal mass
ejection (CME). The chain of events was as follows: (i)~a
presumable eruption at 07:29~UT accompanied by a not reported M1.2
class flare probably associated with the onset of a first
southeastern CME (CME1), which most likely is not responsible for
the superstorm; (ii)~a confined eruption (without a CME) at
07:41~UT (M3.2 flare) that destabilized the large filament;
(iii)~the filament acceleration around 07:56~UT; (iv)~the
bifurcation of the eruptive filament that transformed into a large
``cloud''; (v)~an M3.9 flare in AR 10501 associated to this
transformation. The transformation of the filament could be due to
the interaction of the eruptive filament with the magnetic field
in the neighborhood of a null point, located at a height of about
100~Mm above the complex formed by ARs 10501, 10503, and their
environment. The CORONAS-F/SPIRIT telescope observed the cloud in
304~\AA\ as a large Y-shaped darkening, which moved from the
bifurcation region across the solar disk to the limb. The masses
and kinematics of the cloud and the filament were similar.
Remnants of the filament were not clearly observed in the second
southwestern CME (CME2), previously regarded as a source of the 20
November geomagnetic storm. These facts do not support a simple
scenario, in which the interplanetary magnetic cloud is considered
as a flux rope formed from a structure initially associated with
the pre-eruption filament in AR~10501. Observations suggest a
possible additional eruption above the bifurcation region close to
solar disk center between 08:07 and 08:17~UT that could be the
source of the 20 November superstorm.
\end{abstract}

\end{opening}

\section{Introduction}

\subsection{Enigma of the 20 November 2003 Geomagnetic Superstorm}

Severe geomagnetic storms are among the most dangerous space
weather disturbances caused by solar activity. Impressive examples
are the superstorm on 1\,--\,2 September 1859 produced by the
Carrington event \cite{Carrington1859}, when `many fires were set
up by arching from induced currents in telegraph wires' (E.~Loomis
in \inlinecite{Tsurutani2003}), and the 13\,--\,14 March 1989
superstorm, when a system-wide power blackout occurred in Quebec
(Canada). Strongest geomagnetic storms develop due to the arrival
of magnetic clouds (MC), which originate from solar coronal mass
ejections (CMEs). In a presumable typical scenario, an MC forms
from an erupting magnetic flux rope, which stretches out into
interplanetary space, while its ends usually remain connected to
the Sun. If an Earth-directed MC has a magnetic field component
pointing South ($B_z < 0$), then magnetic reconnection between
such an MC with the Earth's magnetosphere results in a geomagnetic
disturbance. The effect is considered to directly depend on the
magnetic field strength and the MC velocity. Thus, eruptions
involving regions of strongest magnetic fields with the favorable
orientation, \textit{i.e.}, sunspot penumbras and especially
umbras, are major candidates as sources of superstorms
(\textit{cf.} \opencite{Gopal2005c}). Such events produce fast
CMEs, powerful flare emissions up to gamma rays, strong radio
bursts at microwaves, millimeters and even submillimeters, and
energetic particles. Indeed, nine of the 11 superstorms with Dst
$<-250$ nT during Solar Cycle 23 listed by \inlinecite{Echer2008}
were due to major solar eruptive events (GOES X-class flares), but
those on 7 April 2000 and 20 November 2003, the latter being the
strongest one during Solar Cycle 23 (Dst = $-422$ nT according to
the final data of the Kyoto Dst index service,
\url{http://wdc.kugi.kyoto-u.ac.jp/dst_final/index.html}), were
not. These facts imply that either an amplification mechanism
exists or scenarios of such events might be different from a
typical one.

The 20 November 2003 superstorm was probably due to a solar
eruptive event on 18 November. This event occurred during the
second appearance on the Earth-facing solar hemisphere of a
decaying complex of formerly superactive regions (ARs) 10484
(10501), 10486 (10508), and 10488 (10507) (the first numbers
correspond to the previous solar rotation AR numbers. We will use
hereafter the last three digits of the NOAA numbers for brevity)
responsible for the extreme events late in October 2003 (see,
\textit{e.g.}, \opencite{Veselovsky2004};
\opencite{ChertokGrechnev2005}; \citeauthor{Gopal2005a}
\citeyear{Gopal2005a,Gopal2005b}; \opencite{Grechnev2005}).
Several studies (\textit{e.g.}, \opencite{Gopal2005c};
\opencite{Yurchyshyn2005}; \opencite{Ivanov2006};
\opencite{Yermolaev2005}; \opencite{Moestl2008};
\opencite{Srivastava2009}) revealed various features of the 18
November event but have not led to a certain conclusion about the
causes of its extreme geomagnetic effect. The overall scenario of
the event remains unclear. The major question is why this moderate
eruptive event produced an MC with a very strong magnetic field
near Earth of $|B| \approx 56$ nT, extremely large southern
component $B_z \approx -46$ nT, and eventually caused the
superstorm. The success of a recent attempt by
\inlinecite{KumarManoUddin2011} seems to be questionable because
the conjecture about two merged MCs is not supported
quantitatively, while the corresponding large outline in their
Figure~16 disagrees with three reconstructions of the MC
\cite{Yurchyshyn2005,Moestl2008,Lui2011} as all show a compact
cross section. This mysterious event urges one either to find a
key to its extremity or to confess that a possible superstorm can
commence unexpectedly. The only way we see in pursuing the former
option is to analyze the 18--20 November complex event including
eruptions, CMEs, and the interplanetary disturbance in still more
detail. The event has turned out to be the most complex one among
all those we ever dealt with. Its complexity and unusual
properties were misleading for some previous studies. Their
conclusions are compared with the observational facts presented
here. We untangle the complex scenario of the whole event by
following its manifestations and exclude one by one phenomena
whose association with the superstorm is unlikely, as well as the
corresponding conjectures considered previously. On the other
hand, our results obtained in this way shed additional light on
scenarios of eruptions, their course, relations between eruptions
and flares, excitation of shock waves, and other important issues.

We present our study in four companion papers. The present paper
(hereafter paper~I) analyzes eruptions from region 501 on 18
November 2003 and their subsequent evolution. Paper~II (Grechnev
\textit{et al.}, 2013, in preparation) addresses CMEs, coronal
shock waves, and drifting radio bursts, which this event produced.
Paper~III (Uralov \textit{et al.}, 2013, in preparation) addresses
the interaction between the erupting filament and the magnetic
field in the neighborhood of a coronal null point and the helicity
mismatch between the MC and the pre-eruption region. Proceeding
from the results of paper I\,--\,III, paper~IV (Grechnev
\textit{et al.}, 2013, in preparation) tries to understand the
possible causes of the extremity of the 20 November geomagnetic
storm.

\subsection{Challenges of Solar Eruptions}

Solar eruptions are closely associated with flares and CMEs.
Understanding these phenomena is important for both the
fundamental science and space weather forecasting. Despite many
years of studying eruptive phenomena neither persuasive scenarios
of eruptions and CME initiation nor their quantitative
characteristics have been established. Relations between flares,
CMEs, and accompanying disturbances remain under debate. Some
observational challenges do not meet adequate theoretical
expectations. All of these circumstances urge one to analyze
eruptions and associated phenomena in detail, especially
quantitatively.

Eruptive prominences/filaments and surges have been traditionally
observed from Earth in the H$\alpha$ line and more recently in the
He~{\sc i} 10830 \AA\ line. These observations, limited by the
loss of opacity, rarely allow measurements up to long distances.
Space-borne observations in extreme-ultraviolet (EUV) coronal
emission lines of Fe~{\sc ix} 171~\AA\ and Fe~{\sc x--xi} 195~\AA\
carried out by the \textit{Extreme-ultraviolet Imaging Telescope}
(EIT) aboard the \textit{Solar and Heliospheric Observatory}
(SOHO), the \textit{Transition Region and Coronal Explorer}
(TRACE), and recently by the \textit{Extreme UltraViolet Imager}
on the \textit{Solar Terrestrial Relation Observatory} (STEREO)
and the \textit{Atmospheric Imaging Assembly} (AIA) on the
\textit{Solar Dynamics Observatory} (SDO) show filaments in
absorption and sometimes in emission as they are heated. The
He~{\sc ii} 304~\AA\ line is well suited to detect filaments due
to its temperature sensitivity range (maximum at 80\,000~K).
Observations in this line were limited up to the recent time.

A typical eruptive filament (prominence) rises, expands, and
departs from the Sun as a core of a CME. Filament material can
partly drain back to the solar surface along stretched field
lines, former legs of the filament. Sometimes the eruptive
filament stops after the initial rise or returns back to the Sun
(`failed eruptions'). The magnetic structure of an eruptive
filament mainly persists, although partial reconnections seem to
occur both in the filament and between it and the magnetic field
in the environment. Observations also suggest that occasionally an
eruptive filament can transform into a large cloud, which falls
back down to the solar surface far from the eruption region
(\citeauthor{Grechnev2008}, \citeyear{Grechnev2008};
\citeyear{Grechnev2011_AE}). This suggests dramatic changes in the
magnetic structure of the filament. Such little-known anomalous
eruptions have been rarely observed in the 304~\AA\ line.

Most models of eruptions and CME initiation involve magnetic flux
ropes with or without passive filaments embedded and many models
invoke magnetic reconnection. However, the possible role of
filaments themselves in eruptions from active regions is not
highlighted by most models. This might be partially due to
insufficient observational basis of the models because of the
limitations mentioned above and difficulties to observe the
initial phases of CMEs, as well.

The 18 November event manifested in a series of partial eruptions
followed by the eruption of a large filament rooted in AR~501
($\sim\,$N00~E18), a complex long-duration H$\alpha$ flare, and
soft X-ray (SXR) bursts with two reported peaks at 07:52 and 08:31
(\textit{all times hereafter are UT}). The \textit{Large Angle and
Spectroscopic Coronagraph} (LASCO; \opencite{Brueckner1995}) on
SOHO observed two CMEs starting at 08:06 and 08:49. Several papers
addressed various aspects of this solar event, its preparation,
and consequences. Hard X-ray (HXR) emission during the M3.9
two-ribbon flare (after 08:07) in relation to the energy release
rate was studied by \inlinecite{Miklenic2007},
\inlinecite{Moestl2008}, and \inlinecite{Miklenic2009}.
\inlinecite{Srivastava2009} analyzed the evolution of magnetic
fields in AR~501 and compared the magnetic energy budget with
energetics of the flare and CME. \inlinecite{Gopal2005c},
\inlinecite{Yurchyshyn2005}, \inlinecite{Moestl2008},
\inlinecite{Chandra2010}, and \inlinecite{KumarManoUddin2011}
compared properties of the eruptive filament with those of the
magnetic cloud which hit Earth.

The eruptions in this event were observed by several instruments,
but not analyzed in most of the previous papers.
\inlinecite{KumarManoUddin2011} started to study the eruptions,
but their analysis was incomplete, and the discussions were mainly
based on assumptions, although the observations allow one to carry
out detailed measurements.

Our interest in the 18 November 2003 event is reinforced by the
observation of a large inverse-Y-shaped dark feature moving after
the eruption across the solar disk in the South-West direction
towards the limb \cite{Slemzin2004}. This phenomenon was recorded
in 304~\AA\ by the \textit{Spectroheliographic X-ray Imaging
Telescope} (SPIRIT; \inlinecite{Zhitnik2002} and
\inlinecite{Slemzin2005}) aboard the \textit{Complex Orbital
near-Earth Observations of Activity of the Sun} (CORONAS-F)
satellite \cite{{OraevskySobelman2002},{Oraevsky2003}}.
\inlinecite{Grechnev2005} concluded that the darkening was due to
absorption of the solar emission by a cold plasma cloud formed
from an eruptive filament and stated that the filament `probably
failed to become a CME core'. This circumstance has not drawn the
attention of researchers who studied the geoeffective outcome of
the event. From the absorption observed in the He~{\sc ii}
304~\AA\ line, we estimate a density and mass of the absorbing
cloud and compare it with the mass of the eruptive filament. We
also compare their kinematics and analyze coronal magnetic fields.
In this way we endeavor to understand what happened to the
eruptive filament in order to find out in papers II\,--\,IV the
causes of the very strong geoeffective outcome of this eruptive
event.

Section~\ref{S-description} outlines observations of the solar
event and reveals its important episodes.
Section~\ref{S-kinematics} addresses measurements of the
kinematics. Section~\ref{S-magnetic_fields} studies magnetic
fields. In Section~\ref{S-masses} we compare the masses of the
eruptive filament and the Y-like cloud observed in 304~\AA. We
summarize and discuss our results in Section~\ref{S-discussion}.
Section~\ref{S-conclusion} presents the major outcome of the
study.

\section{Observations}
\label{S-description}

\subsection{Instruments and data}

The event was observed in Kanzelh{\"o}he Solar Observatory (KSO)
with a tunable Lyot H$\alpha$ filter (full width at half-magnitude
(FWHM) of 0.7~\AA) mostly in the center of the 6562.8~\AA\
H$\alpha$ line (line-of-sight velocities $V_{\mathrm{LOS}} = \pm
16$~km~s$^{-1}$, the sunward direction is positive), in the blue
wing [$-0.365$~\AA, $V_{\mathrm{LOS}} = -(0.6-33)$~km~s$^{-1}$],
and in the red wing [$+0.442$~\AA, $V_{\mathrm{LOS}} =
+(4-36)$~km~s$^{-1}$]. On average, six line-center images and one
pair in the wings were produced every minute from 07:30:32 to
09:20:18.

We use images observed with TRACE \cite{Handy1999} in the 171~\AA\
and 195~\AA\ channels, which respond to normal coronal
temperatures. There was a gap in the observations between 07:00:26
and 07:33:17. The field of view centered at $\sim [-200^{\prime
\prime}, 0^{\prime \prime}]$ was $1024 \times 1024$ in the
171~\AA\ channel and $768 \times 768$ in the 195~\AA\ channel with
a $0.5^{\prime \prime}$ pixel size.

CORONAS-F/SPIRIT observed simultaneously in the He~{\sc ii}
304~\AA\ band and the Fe~{\sc ix\,--\,xi} 175~\AA\ coronal band
(predominant temperature of 1~MK). SPIRIT produced one pair of
images in both bands typically every 15 min which determines its
opportunity to study eruptions (methods of data processing can be
found in \inlinecite{Bogachev2009}). We also use images of
SOHO/EIT \cite{Delab1995} and the \textit{Soft X-ray Imager} (SXI;
\opencite{Hill2005}; \opencite{Pizzo2005}) on GOES-12, produced
with a polyimide filter composition. A wide SXI passband, with a
maximum at $\approx 4$~MK, is sensitive to temperatures of $\gsim
10$~MK at a level of 20\%. The sensitivity also declines to 10\%
at $\approx 1$~MK.

Magnetic fields were studied from full-disk magnetograms produced
with the \textit{Michelson Doppler Imager} (MDI;
\opencite{Scherrer1995}) on SOHO. We study the evolution of
magnetic fields on the photosphere from 1-min series of MDI
magnetograms produced from 07:00 to 08:00. To analyze the coronal
magnetic field, we used for extrapolations three MDI magnetograms
from 96-min series observed at 06:23, 07:59, and 09:36. In spite
of active photospheric motions, the coronal configuration of
interest did not show significant changes. The results presented
here have been obtained from the MDI magnetogram at 07:59.

\subsection{A Brief Description of the Event}

\begin{figure} % {1}
  \centerline{\includegraphics[width=\textwidth]
   {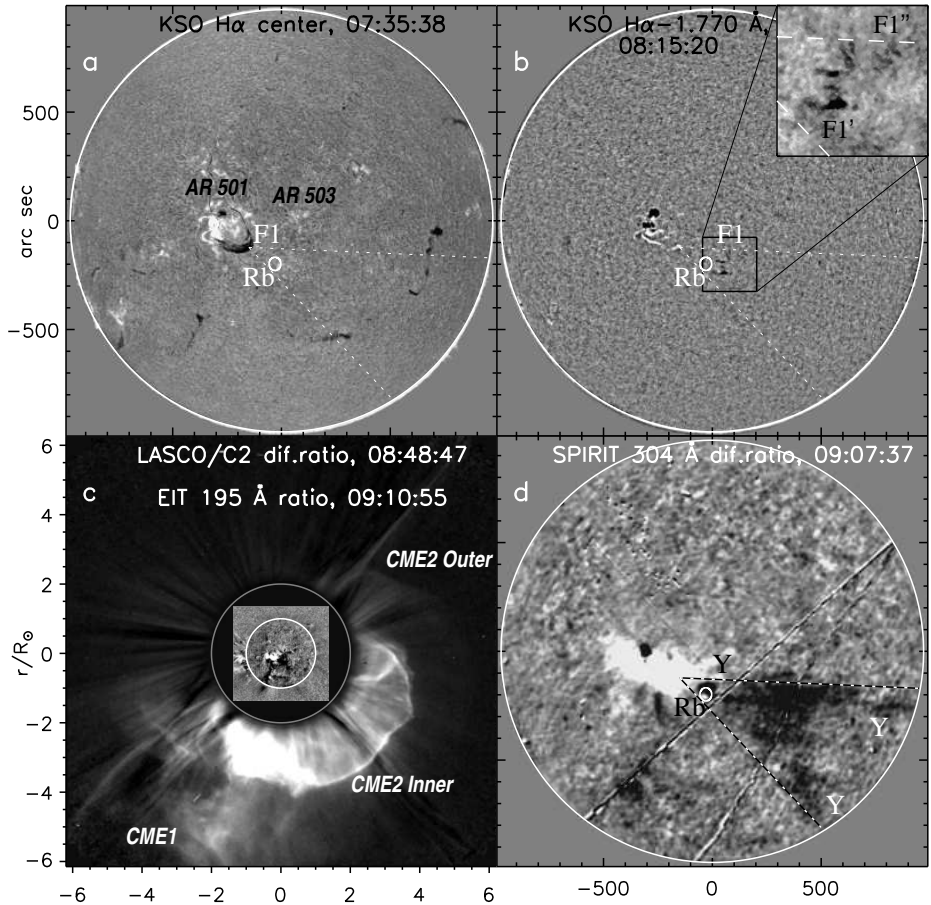}
  }
  \caption{Overview of the event. a)~The main filament F1
before eruption (H$\alpha$, KSO). A flare in the eastern part of
AR~501 already started. b)~The eruptive filament in the far blue
wing H$\alpha$ image. Flare ribbons expanded west. c)~CMEs
observed with LASCO/C2 and their footprints in a later EIT
195~\AA\ image ratio. d)~A large Y-like darkening in a
CORONAS-F/SPIRIT 304~\AA\ image. The broken lines in panels a),
b), and d) outline a sector in which the eruptive filament and the
Y-like darkening were observed. The oval labeled `Rb' denotes the
region of bifurcation. The axes show distances from the solar disk
center in arc seconds (panels a), b), and d)) and in solar radii
(panel c)). }
  \label{F-overview}
  \end{figure}

Active region 501 (Carrington longitude $L_0=2$) exhibited ongoing
activity. An M4.2 SXR class and 1N H$\alpha$ flare, associated to
a CME event, on 17 November (SXR peak at 09:05) and a series of
events on 18 November occurred in this region. The main 18
November event was associated with the eruption of a large
U-shaped filament (F1) that extended to the South-West of AR~501
(Figure~\ref{F-overview}a). The main eruption (after 07:55) was
preceded by partial eruptions of some portions of the filament.
Figure~\ref{F-overview}b reveals the eruptive filament in the far
blue wing of the H$\alpha$ line, at least, 15 min later than the
onset of its eruption. The filament looks reversed and fragmented
in the inset.

A LASCO/C2 image in Figure~\ref{F-overview}c shows CME1, probably
associated with one of partial eruptions, and CME2 presumedly
ejected during the main eruption (the onset time of CME2
extrapolated to the solar disk center was between 08:05 and 08:20
\cite{Gopal2005c}). CME2 had a faster, fainter outer halo (speed
$\approx\,$1660 km~s$^{-1}$) and a brighter, slower (by about
30\%) inner component. An arcade and dimmings in a later EIT
195~\AA\ image ratio (09:11 divided by an image at 07:35) suggest
eruptions and possibly other strong perturbations in the corona.
CORONAS-F/SPIRIT observed a Y-like darkening in 304~\AA\
(Figure~\ref{F-overview}d) well after the eruption and appearance
of CME2.

\subsection{X-ray and Microwave Time Profiles}

\begin{figure} % {2}
  \centerline{\includegraphics[width=\textwidth]
   {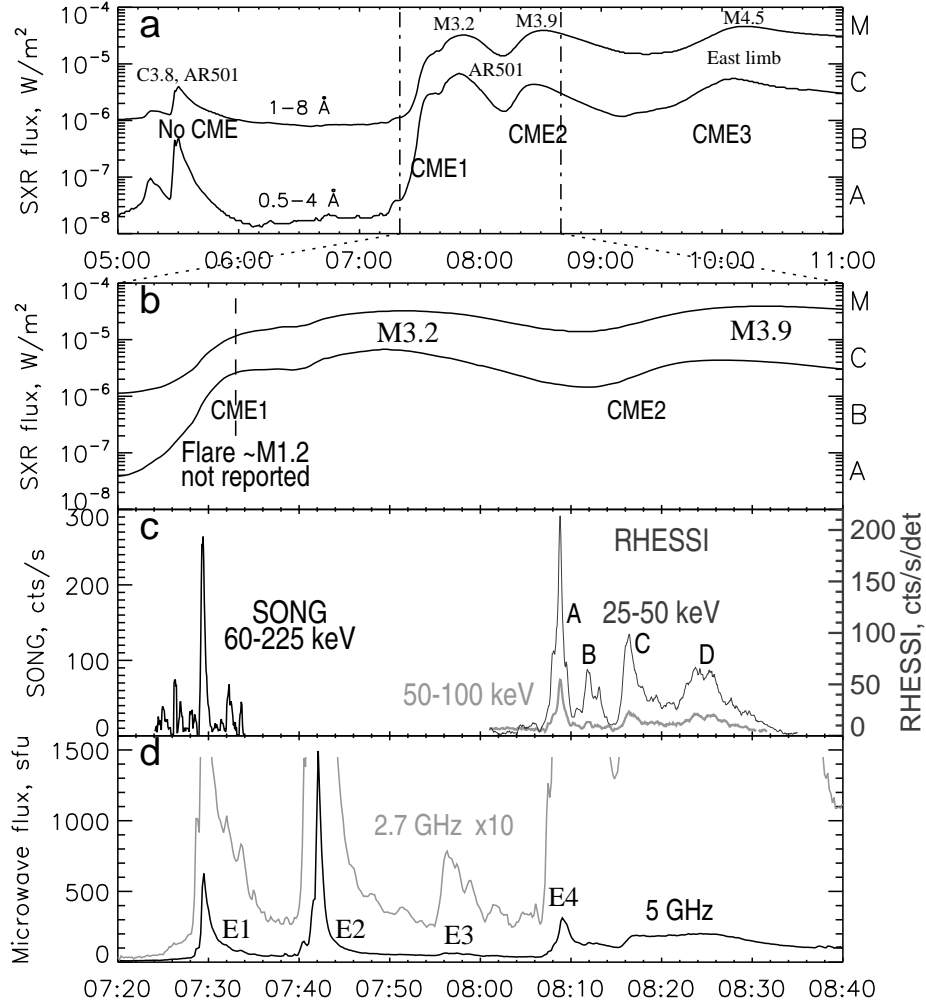}
  }
  \caption{a) and b) GOES SXR flux, b)~shows in more detail the interval
between vertical dashed lines in a). c)~HXR flux, the gray curve
corresponds to the higher energy range. d)~Microwave light curves,
the gray curve in this panel shows a 2.7 GHz time profile
multiplied by a factor of 10.}
  \label{F-goes_rstn}
  \end{figure}

Figure~\ref{F-goes_rstn}a presents SXR GOES light curves for a
six-hour interval on 18 November. A C3.8 eruptive flare occurred
in AR~501 from 05:00 to 06:00 and the series of events of
interest, shown in the lower panels of Figure~\ref{F-goes_rstn},
started at 07:14 with a group of type III bursts. A corresponding
2N H$\alpha$ flare was reported to occur from 07:16 to 09:55,
extending from S01~E20 to N00~E18. The initial episode was weak,
with an SXR class of $<$~C1.3 level, and was not a conspicuous
event. A later eruptive event, with an M4.5 SXR peak at 10:11,
occurred at the East limb in appearing AR~508 (returning AR~486).
This limb event was related to onset of a third CME, CME3
\cite{Gopal2005c}, which could not reach Earth.

Figure~\ref{F-goes_rstn} shows the time profiles of the activity
just described from 07:20 to 08:40 in SXR (b), HXR (c), and
microwaves (d). Some of the episodes were not studied previously.
GOES reported two SXR bursts of an M3.2 flare (peak at 07:52) and
an M3.9 flare (peak at 08:31). The \textit{Solar Neutrons and
Gamma} multi-channel detector (SONG; \opencite{Kuznetsov2011})
aboard CORONAS-F registered a short ($\sim\,$1~min) HXR burst at
07:29. The \textit{Reuven Ramaty High-Energy Solar Spectroscopic
Imager} (RHESSI; \opencite{Lin2002}) recorded a longer complex HXR
burst starting at 08:07. No HXR data are available between 07:35
and 08:00 due to night-time intervals of both of instruments, and,
therefore, we use microwaves as a proxy of hard X-rays. Microwave
light curves in Figure~\ref{F-goes_rstn}d show a series of bursts
recorded at 5 and 2.7 GHz in San Vito (US Air Force Radio Solar
Telescope Network). The whole event consisted of four consecutive
episodes, E1\,--\,E4. Their association with CMEs corresponds
approximately to the identification of \inlinecite{Gopal2005c} and
will be confirmed in this paper and paper~II.

The SXR flux of E1 reached a level of $\sim\,$M1.2. This SXR burst
overlapped with the next one (M3.2 at 07:52); the SXR flux had not
decreased between the two bursts and the M1.2 event was not
reported. TRACE missed the corresponding eruption and we are not
aware of its presence in H$\alpha$ observations. The presumable
eruption at that time is confirmed by the appearance of the
southeastern CME1. Episode E2 during RHESSI and CORONAS-F
night-time was strongest in microwaves (07:42) and reached an M3.2
level. A faint episode E3 at $\sim\,$07:57 is barely detectable in
microwaves, but appears to be related to an important eruption.
Episode E4 was weaker than E1 and E2, but longer
(08:07\,--\,08:32), with a highest peak E4A at 08:09. The fourfold
HXR burst E4 was studied in detail by \inlinecite{Miklenic2007}
and \inlinecite{Moestl2008}. The peaks of E4 are denoted A, B, C,
and D according to the former paper. We show two HXR bands in
Figure~\ref{F-goes_rstn}c. Higher-energy hard X-rays are known to
match microwaves better; indeed, the 50\,--\,100 keV time profile
is rather similar to the 5~GHz flux. Thus, microwaves can be used
as a proxy of missing HXR data, but peak E4B was faint in
microwaves.

To summarize, due to the intrinsically gradual character of the
SXR emission (see, \textit{e.g.}, \opencite{Neupert1968}), event
E1 was not registered in soft X-rays. The SXR response to a series
of HXR bursts after 08:07 merged into a single M3.9 burst. Episode
E3 was too weak to be detected. Our analysis has revealed the
following presumable chain of events and associations with
E1\,--\,E4 (the measured eruptions started $\approx 1$~min before
the peaks in HXR or microwaves):
\begin{itemize}
 \item
07:29, E1. Probable eruption and CME1 onset. GOES M1.2 level.

 \item
07:41, E2. Confined ejection. No CME. The U-shaped filament
departed. GOES M3.2 level.

 \item
07:56, E3. Stronger acceleration of the U-shaped filament.

 \item
08:09, E4A. Collision of the eruptive filament with an
``obstacle'' close to the solar disk center.

 \item
08:12, E4B. Bifurcation of the eruptive filament.

 \item
08:16, E4C. Presumable additional eruption above the region of
bifurcation.

 \item
08:22--08:32, E4D. To be studied in paper II. GOES M3.9 level.

 \item
08:23--09:55. Moving Y-like cloud formed by filament remnants.
\end{itemize}

\subsection{Partial Filament Eruptions}

The U-shaped filament (or a filament system) F1 was active, except
for its thin north part. An eruption in its eastern part on 17
November at about 09:00 produced an M4.2/1N flare, a jet-like
ejection to the West, and a southeastwards CME with a central
position angle (PA) of about $135^\circ$. Its geomagnetic effect
was not pronounced. The main body of the U-shaped filament
remained static.

  \begin{figure} % {3}
  \centerline{\includegraphics[width=\textwidth]
   {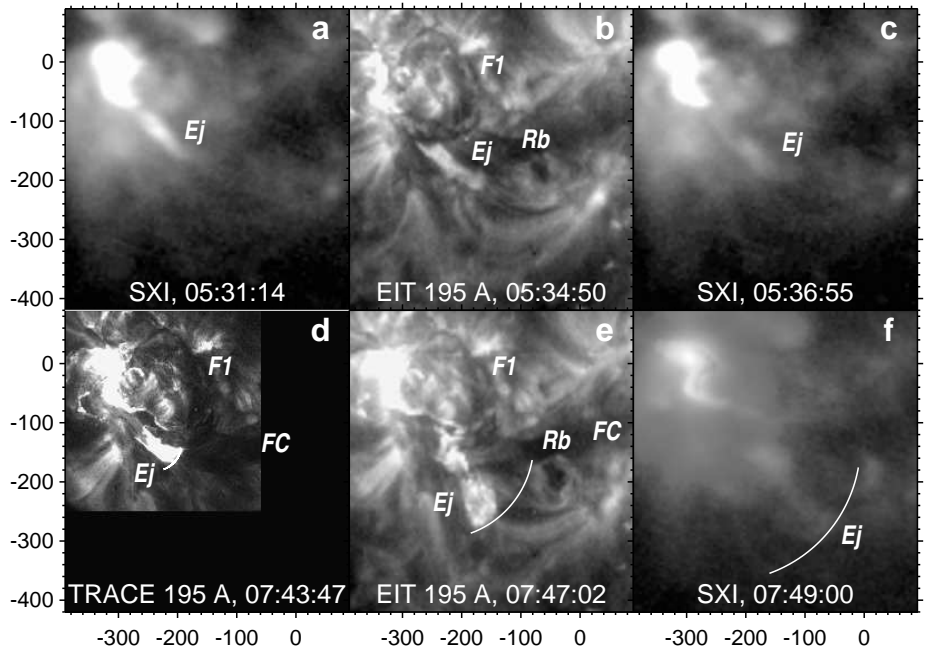}
  }
  \caption{Homologous eruptions on 18 November:
a\,--\,c)~05:30 event (C3.8 flare), d\,--\,f)~07:41 event (M3.2
flare). F1 is the main U-shaped filament, Ej are bright ejecta, Rb
is the bifurcation region, FC is a static filament channel. The
arc in the lower panels corresponds to the fit in
Figure~\ref{F-eruption2_fit}. The axes show hereafter the
distances from the solar disk center in arc seconds.}
  \label{F-homol_eruptions}
  \end{figure}

On November 18, an eruption related to the C3.8/SF flare at 05:30
occurred also in the eastern part of AR~501 (S01E20,
Figures~\ref{F-homol_eruptions}a\,--\,c). Ejecta, called Ej in
Figure~\ref{F-homol_eruptions}, moved along the loops with ends
rooted to the West of region Rb. The eruption did not originate
any CME conversely to what was proposed by
\inlinecite{KumarManoUddin2011}, who related this flare with a
slow CME which appeared at 05:26. The SOHO/LASCO CME catalog
(\url{http://cdaw.gsfc.nasa.gov/CME_list}, \opencite{Yashiro2004})
estimates its average speed to be 267~km~s$^{-1}$ and linearly
extrapolates the onset time to 04:13 at $1R_{\odot}$ and to 03:29
at the disk center (close to AR~501). According to
\inlinecite{Zhang2001} and \citeauthor{Temmer2008}
(\citeyear{Temmer2008,Temmer2010}), the largest acceleration of a
CME is temporally close to a related flare. The large time
difference between the C3.8 flare and the onset time of this CME
rules out their association.

The ejection associated with E2 at 07:41
(Figures~\ref{F-homol_eruptions}d\,--\,f) was similar to the
preceding eruptions. A bright ejecta moved initially to the
South-West, then turned south, and after that turned west again
following the way of the eruption at 05:30 traced by coronal
loops.

\begin{figure} % {4}
  \centerline{\includegraphics[width=\textwidth]
   {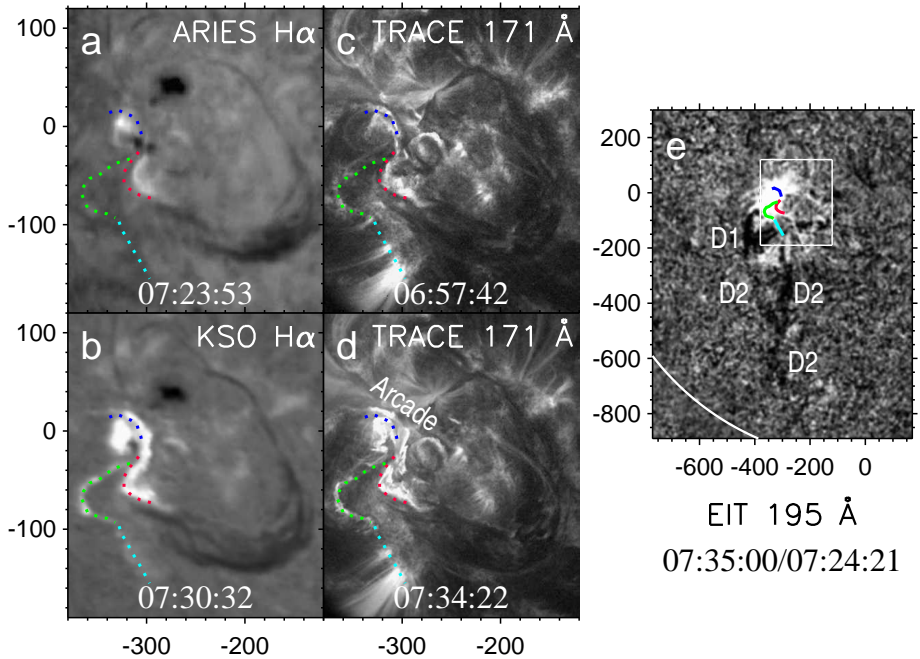}
  }
  \caption{The not reported M1.2 eruptive event at about
07:29. a,\,b)~Development of the flare in ARIES and KSO H$\alpha$
images. c,\,d)~TRACE 171~\AA\ images before and after the
eruption, respectively, a post-eruption arcade is visible in c).
e)~Flare and dimmings (D1 and D2) in a post-eruption EIT image
ratio. The color lines outline the flare brightenings and
preceding structures. The frame in panel (e) shows the field of
view in panels a\,--\,d.}
  \label{F-event_0730}
  \end{figure}

The unreported episode E1 at 07:29 occurred also in the eastern
part of AR501. Several facts suggest a related eruption. The
images in Figures~\ref{F-event_0730}a,\,b indicate the development
of an H$\alpha$ flare (the image produced at Aryabhatta Research
Institute of Observational Sciences (ARIES) is from the paper by
\opencite{KumarManoUddin2011}). The color lines trace the flare
ribbons and preceding structures. A small flare arcade appeared in
171~\AA\ between 06:57 and 07:34
(Figures~\ref{F-event_0730}c,\,d). The EIT 195~\AA\ image ratio in
Figure~\ref{F-event_0730}e reveals the large-scale character of
the eruption. Dimming D1 was probably due to the displacement of
some long loops and eruption of some others. A complex dimming D2
extending far south suggests the eruption of large structures.
According to \inlinecite{Gopal2005c}, the onset time of CME1 at PA
$\approx 168^{\circ}$ extrapolated to the solar disk center was
about 07:22, closer to the 07:29 event than to the 07:41 eruption.
Our measurements in paper~II confirm this timing.

In summary, the homologous partial jet-like eruptions either had
not produced any CME at all (18 November, 05:30 and 07:41) or were
not geoeffective (17 November). A geomagnetic effect of the poorly
observed eruption at 07:29, presumedly responsible for CME1 onset,
is doubtful, as discussed in paper~II. The main part of the
filament survived after all of these eruptions and fully erupted
only after 07:55. These circumstances imply that the geomagnetic
storm was most likely related to the eruption of the main filament
rather than any of the preceding partial eruptions. Nevertheless,
the eruption at 07:41 deserves attention, because it could
destabilize the main part of the U-shaped filament and produce
some observed disturbances.

\subsubsection{Eruption of a Bright Jet at 07:41 (E2)}

\begin{figure} % {5}
  \centerline{\includegraphics[width=\textwidth]
   {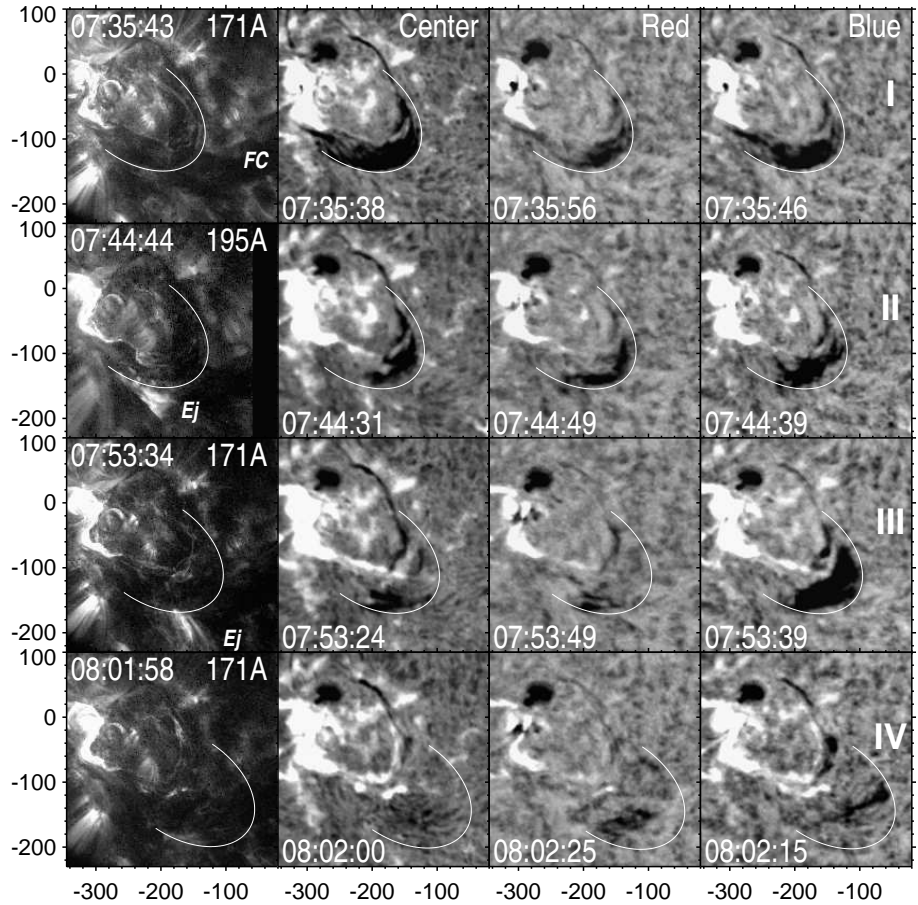}
  }
  \caption{Eruptions E2 and E3 in TRACE and
H$\alpha$ images. The white arc outlines the darkest part of the
filament according to the solid fit in
Figure~\ref{F-eruption3_plots}. The shape and size of the arc
correspond to the initial filament. FC is a static filament
channel. Ej is a bright ejection visible in the TRACE images.}
  \label{F-ha_images}
  \end{figure}

The EUV and H$\alpha$ images in Figure~\ref{F-ha_images} show the
eruptions after 07:30 (see also the movies TRACE.mpg,
SPIRIT304.mpg, and SXI.mpg in the electronic version of the
paper). The main filament initially looked similar in EUV,
H$\alpha$ line center, and both H$\alpha$ wings, probably, due to
the presence of internal plasma flows. A bright jet-like feature,
Ej, appeared after 07:41 in the South-East leg of the filament and
moved initially along it (left column, second and third rows).
Then it transformed into a bright untwisting bundle of threads and
turned South. Its subsequent motion is visible in the EIT and
GOES/SXI images in Figures~\ref{F-homol_eruptions}e,\,f.

When the ejecta appeared, the corresponding portion of the
filament vanished. After that the remaining part of the filament
was observed to be irregular and different in H$\alpha$ line
center and both wings. However, the filament persisted to be
steady in TRACE images and only started to move slowly to the
South-West. These facts suggest a perturbation of plasma flows in
the filament as well as heating effects, both induced by the
jet-like eruption.

\subsubsection{Eruption of the Main Filament at 07:56 (E3)}

\begin{figure} % {6}
  \centerline{\includegraphics[width=\textwidth]
   {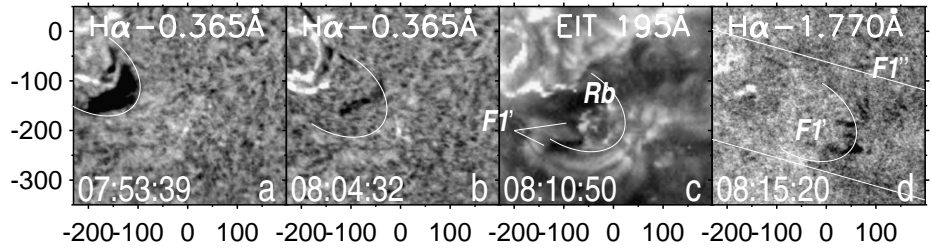}
  }
  \caption{Motion of the eruptive filament in
the blue wing of H$\alpha$ and EUV images. Rb is the region of
bifurcation. Two segments starting from F1$^{\prime}$ label
delimit the portion of filament F1 visible in the EIT 195~\AA\
image c). A wide dark inclined band in this image is a static
structure. F1$^{\prime}$ and F1$^{\prime \prime}$ are two parts of
the filament (see Figure~\ref{F-bifurcation}). The white arcs
correspond to the fit in Figure~\ref{F-eruption3_plots}. The white
lines delimit the band used for the measurements presented in
Figure~\ref{F-eruption3_red}.}
  \label{F-eruption3_images}
  \end{figure}

Filament F1 underwent major acceleration around 07:56 and moved to
the South--West losing opacity. Figure~\ref{F-eruption3_images}
shows its motion in the blue H$\alpha$ wing and an EIT 195~\AA\
image. Our description is related henceforth to projected
manifestations of phenomena which occurred at heights $\sim
100$~Mm. The projected trajectory of the filament crossed region
Rb. Having approached this region, the eruptive filament became
concave-outward in Figure~\ref{F-eruption3_images}c and then
bifurcated into two parts F1$^{\prime}$ and F1$^{\prime \prime}$,
which moved around region Rb as if this region hampered the motion
of the filament's middle part. After the ``collision'' with region
Rb, the eruptive filament looks fragmented, reverted, shifting
North in the later far-blue-wing image in
Figure~\ref{F-eruption3_images}d, while F1$^{\prime \prime}$ moved
faster than F1$^{\prime}$.

The southern filament part F1$^{\prime}$ is only visible in
Figure~\ref{F-eruption3_images}c against bright coronal loops,
while part F1$^{\prime \prime}$ is indistinguishable against a
dark filament channel underneath. This static channel was
separated by a wide unipolar gap from the photospheric inversion
line related to the pre-eruptive filament F1. This filament
channel was passive during the event. This fact is significant to
identify a source structure for the 20 November magnetic cloud
(paper~III).

  \begin{figure} % {7}
  \centerline{\includegraphics[width=\textwidth]
   {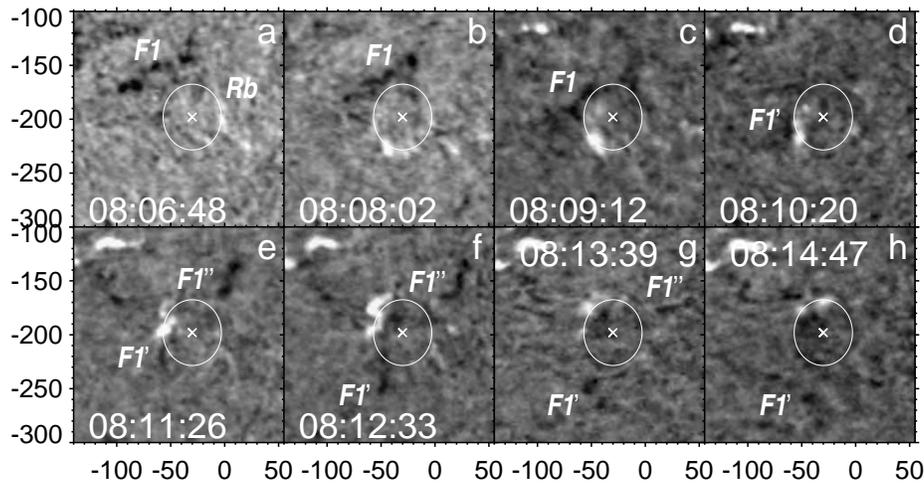}
  }
  \caption{Bifurcation of the filament and a rotating motion underneath
in the blue-wing H$\alpha$ images. F1 is the eruptive filament,
F1$^{\prime}$ and F1$^{\prime \prime}$ are its parts after the
collision with region of bifurcation Rb indicated with an oval.
The slanted crosses mark the center of Rb.}
  \label{F-bifurcation}
  \end{figure}

\begin{figure} % {8}
  \centerline{\includegraphics[width=\textwidth]
   {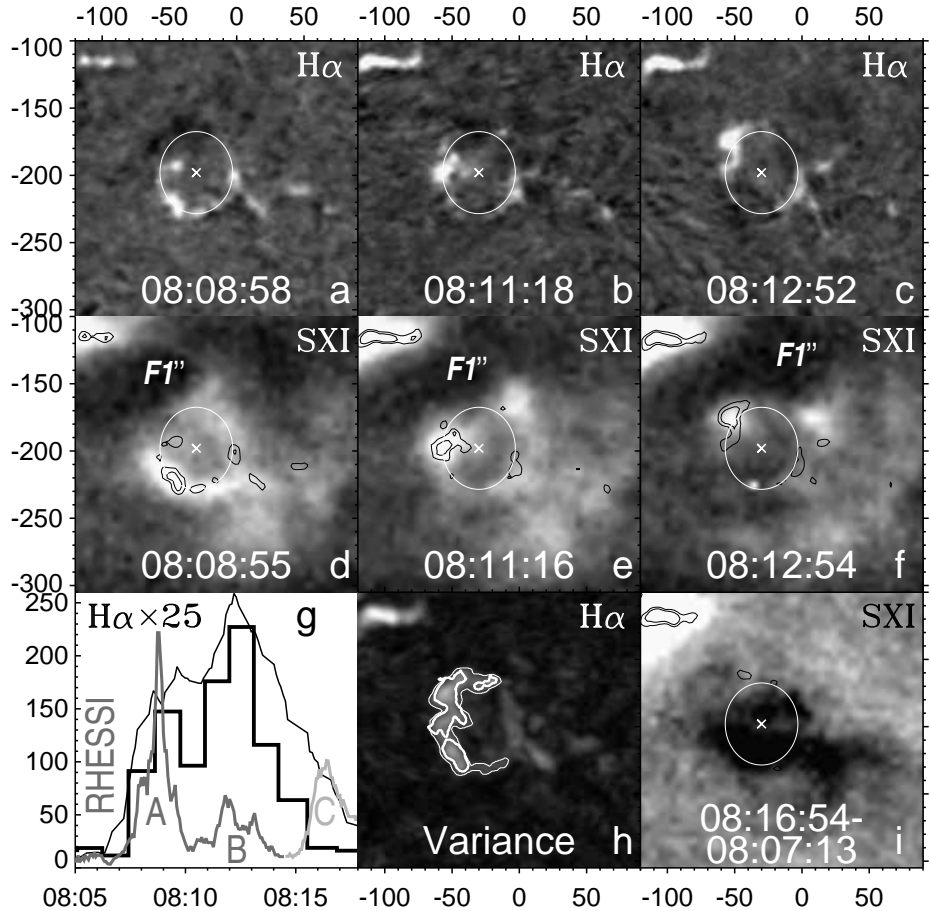}
  }
  \caption{Flare-like phenomena in the bifurcation region in
H$\alpha$ (a\,--\,c) and GOES/SXI (d\,--\,f) images. The contours
on top of the SXI images outline the H$\alpha$ brightenings
(levels of 1/8 and 1/4 of the maximum brightness). g)~Time
profiles in the H$\alpha$ line center (black thin solid line) and
the blue wing (thick histogram) computed for the regions contoured
in panel (h) with the corresponding thicknesses. The gray curve
shows the RHESSI time profile in 25\,--\,50 keV. All the time
profiles are quantified in instrumental units. h)~Variance map
computed from the line-center H$\alpha$ images. i)~Dimming in an
SXI difference image.}
  \label{F-ha_sxi}
  \end{figure}

Figure~\ref{F-bifurcation} presents the response in the H$\alpha$
line to the collision, which occurred high in the corona above
region Rb. Brightenings appeared in the southern part of this oval
at about 08:07, when the approaching filament apparently contacted
region Rb, and then moved clockwise by about 08:17 with a total
rotation angle of $\sim 180^{\circ}$. The motion in region Rb is
visible in both, the H$\alpha$ line center and the blue wing, but
is not pronounced in the red wing; this indicates prevailing
upward plasma motions. GOES/SXI images in Figure~\ref{F-ha_sxi}
also show the rotating brightenings. The coronal brightenings were
more diffuse than the H$\alpha$ ones, but both coincided
spatially, as the contours of the H$\alpha$ brightenings on top of
the SXI images show. The SXI images show the northern filament
part F1$^{\prime \prime}$ to roll around the bifurcation region Rb
(\textit{cf.} Figures \ref{F-bifurcation}c,\,e,\,f and Figures
\ref{F-ha_sxi}d,\,e,\,f). The dominant plasma temperatures in the
coronal brightenings probably exceeded 1~MK, as suggested by the
SXI temperature sensitivity range. Presumably, low coronal layers
responded to the interaction between magnetic structures of the
filament, which moved high above the photosphere, with static
coronal structures rooted in region Rb. The filament is not
apparent in H$\alpha$ images after 08:16.

The rotating features were brightest at about 08:09 and 08:12,
when the HXR peaks E4A and E4B occurred. Figure~\ref{F-ha_sxi}g
shows H$\alpha$ light curves computed for a variable part of
region Rb contoured in Figure~\ref{F-ha_sxi}h (the variance
analysis was described by \inlinecite{Grechnev2003}) along with
the HXR burst (gray). The thin black curve is the H$\alpha$
line-center time profile computed for the region within the thin
contour in Figure~\ref{F-ha_sxi}h. The enhancements in the
H$\alpha$ time profile correspond to the HXR peaks E4A and E4B.
The correspondence is still more pronounced in the thick black
time profile computed for the region within the thick contour in
Figure~\ref{F-ha_sxi}h from blue-wing images observed with a lower
rate. A later HXR peak E4C (light gray) had no counterpart in the
bifurcation region suggesting its disconnection from AR~501
between E4B and E4C.

A difference SXI image in Figure~\ref{F-ha_sxi}i shows a dimming
region which developed around region Rb from 08:11 to 08:17. This
fact and the disappearance of the brightenings by 08:15
corresponding to the absence of peak E4C in the H$\alpha$ light
curve in Figure~\ref{F-ha_sxi}g, suggests a possible additional
eruption above the bifurcation region.

\subsection{Disturbances Observed in 175 and 304~\AA}

\begin{figure} % {9}
  \centerline{\includegraphics[width=\textwidth]
   {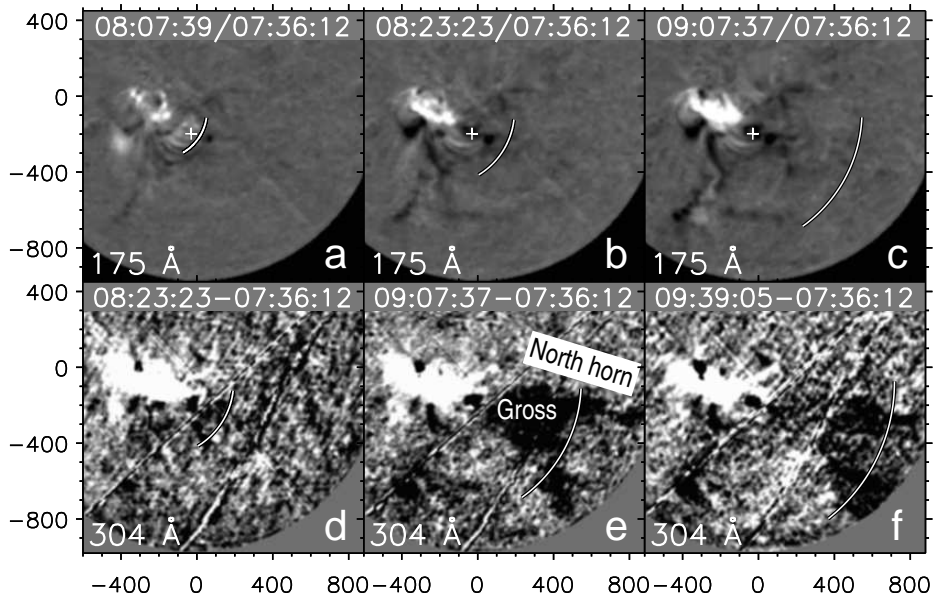}
  }
  \caption{CORONAS-F/SPIRIT observations. The upper
row shows 175~\AA\ image ratios for the times indicated at the top
of each panel. The white cross denotes the center of the
bifurcation region, Rb. The lower row presents a moving Y-like
darkening in enhanced-contrast 304~\AA\ difference images. The
arcs correspond to the fit shown in
Figure~\ref{F-eruption3_plots}. Slanted fracture-like features are
due to instrumental effects.}
  \label{F-spirit}
  \end{figure}

Figure~\ref{F-spirit} shows CORONAS-F/SPIRIT observations. The
upper row presents 175~\AA\ image ratios. A brightening around the
cross to the South-West of AR~501 in Figure~\ref{F-spirit}a
corresponds to the position of the eruptive filament at that time.
The brightening could be due to displacement and compression of
coronal loops by the expanding filament. The 175~\AA\ images show
the development of a complex dimming. The central dimming region
around the cross corresponds to the dimming in a GOES/SXI image
(Figure~\ref{F-ha_sxi}i), supporting the occurrence of an eruption
at that place.

The 304~\AA\ difference images in the lower row reveal a large
dark inverted-Y-like feature moving to the South-West. It is
conspicuous in Figure~\ref{F-spirit}e (09:07) but appeared
possibly earlier; the arc in Figure~\ref{F-spirit}d hinting at its
expected position at 08:23 allows one to detect a faint indication
of a Y-shaped feature. The directions of motion of the eruptive
filament and the Y-like feature were close to each other. No
counterparts of the Y-darkening are visible in the 175~\AA\
images.

The changes in the shape of the Y-like feature in
Figures~\ref{F-spirit}d\,--\,f resemble a shrinkage because of
projection in a motion along a nearly spherical surface
constituted by very long closed magnetic field lines. These
circumstances hint at a probable association of the Y-like
darkening with the eruptive filament and its failure to leave the
Sun. To verify this suggestion, we will compare the kinematics of
the filament and the Y-darkening, as well as their estimated
masses.

\section{Kinematics}
\label{S-kinematics}

\subsection{Method}
 \label{S-kinematics_method}

Several studies (\textit{e.g.}, \opencite{Zhang2001};
\opencite{Maricic2007}; \citeauthor{Temmer2008},
\citeyear{Temmer2008,Temmer2010}) conclude that the acceleration
of a CME agrees in time with the appearance of an associated HXR
burst. However, this does not guarantee the same correspondence
for eruptive filaments. \inlinecite{Grechnev2011_I} measured
accelerations of small eruptive magnetic ropes in two events to be
temporally close to HXR/microwave bursts, but not to coincide with
them perfectly.

The acceleration time profile is usually inferred by
differentiation of distance-time measurements. This way is
critical to measurement errors, temporal sampling, and provides
large uncertainties. Even sophisticated methods (\textit{e.g.},
\opencite{Maricic2004}; \opencite{Temmer2010}) do not overcome
these difficulties completely. Our approach based on fitting an
analytic function to measurements \cite{Grechnev2011_I} is based
on the fact that the initial and final velocities of an eruption
are nearly constant and acceleration occurs within a rather short
time interval (see, \textit{e.g.}, Figure 1b in
\opencite{Rompolt1998}). We describe the acceleration with a
Gaussian time profile (see \opencite{WangZhangShen2009}), $ a =
\left( v_1-v_0 \right) \exp{\{-{[(t-t_0)/\tau_{\mathrm{acc}}]^2}/2
\}} /
  (\sqrt{2\pi}\tau_{\mathrm{acc}})$.
Here $\tau_{\mathrm{acc}}\sqrt{8\ln{2}}$ is the FWHM of the
acceleration time profile centered at time $t_0$ and $v_0$ and
$v_1$ are the initial and final velocities. Kinematical plots are
calculated by means of integration or differentiation of the
analytic fit rather than the measurements. Our ultimate criterion
is to follow the motion of an analyzed feature in images as
closely as possible. In cases of more complex kinematics, we use a
combination of Gaussians and adjust their parameters manually. The
main uncertainties of the measurements are due to lamination of a
filament into a multitude of expanding faint threadlike fragments
that complicates following the same moving feature.

\subsection{Impulsive Eruption at 07:41 (E2)}

\begin{figure} % {10}
  \centerline{\includegraphics[width=\textwidth]
   {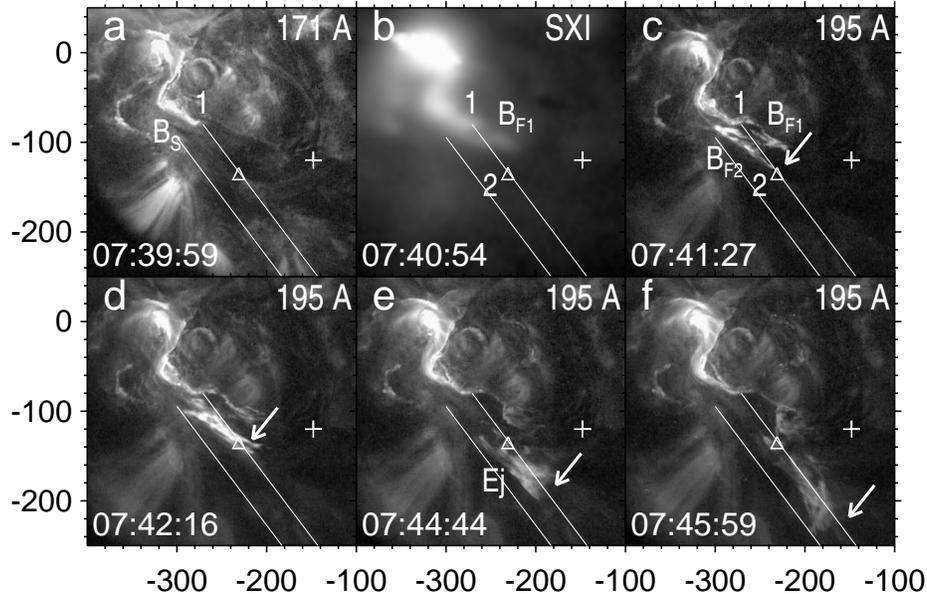}
  }
  \caption{Impulsive eruption E2 in TRACE 171 and 195~\AA\
and GOES/SXI images (the latter is shown in b)). The triangle is
the origin of measurements. The cross marks the initial position
of the filament's top. The Solar rotation is compensated to 08:00.
The arrows point to the leading edge corresponding to the fit in
Figure~\ref{F-eruption2_fit}. The white lines delimit a slice used
in the measurements.}
  \label{F-eruption2_images}
  \end{figure}

\begin{figure} % {11}
  \centerline{\includegraphics[width=\textwidth]
   {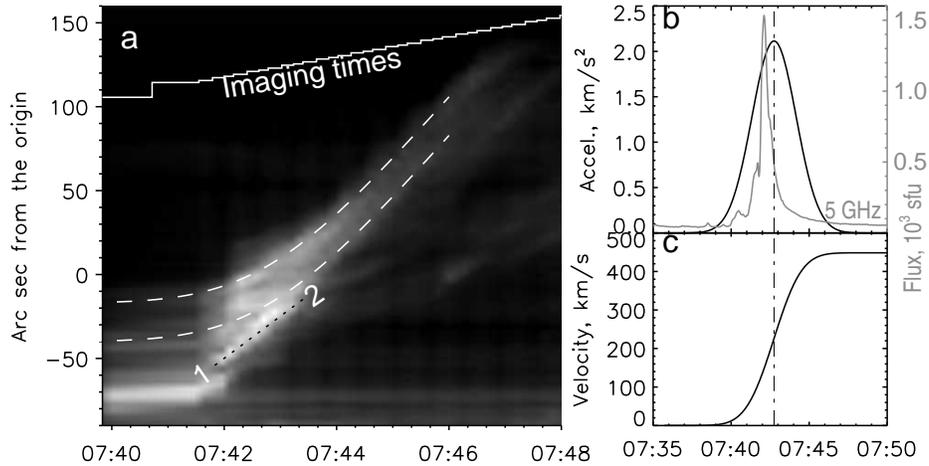}
  }
  \caption{Plane-of-sky kinematics for ejection E2.
(a)~Spatial profiles computed from TRACE 171 and 195~\AA\ images.
The upper histogram-like plot shows the sampling intervals. The
black dotted line 1\,--\,2 marks the feature with a sharpest
apparent acceleration. (b,c)~Acceleration and velocity profiles of
the ejection corresponding to the white dashed curves in panel
(a). The gray curve in panel (b) shows the microwave burst at 5
GHz.}
  \label{F-eruption2_fit}
  \end{figure}

TRACE observed the eruption in the 171 and 195~\AA\ channels
(Figure~\ref{F-eruption2_images} and TRACE.mpg and SXI.mpg
movies). The initial imaging rate was relatively low (upper
histogram-like plot in Figure~\ref{F-eruption2_fit}a). To reduce
the time interval between the two TRACE images in
Figures~\ref{F-eruption2_images}a and \ref{F-eruption2_images}c,
we use a GOES/SXI image in Figure~\ref{F-eruption2_images}b. The
first TRACE image shows the pre-eruptive filament. The dark
filament blocked the observation of the structures behind it. The
large opacity of the dark filament material indicates its low
temperature, $< 2\times 10^4$~K. There was also a bright
mirrored-S-shaped feature, B$_\mathrm{S}$, whose northwestern end
is labeled as 1.

Portion B$_\mathrm{F1}$ of the filament South-West from 1
brightened up in Figures~\ref{F-eruption2_images}b,\,c indicating
a temperature $\gsim 1$~MK, according to the sensitivity ranges of
the TRACE 195~\AA\ channel and GOES/SXI. The average plane-of-sky
speed of the brightening between Figures \ref{F-eruption2_images}a
and \ref{F-eruption2_images}b was $\geq 550$~km~s$^{-1}$. Motions
of B$_\mathrm{F1}$ were much slower later on. These facts suggest
that the rapid brightening B$_\mathrm{F1}$ over a distance of
$\approx 30$~Mm was most likely ignited by an MHD disturbance.

Starting from 07:41:27 (Figure~\ref{F-eruption2_images}c), the
TRACE images also show another brightening B$_\mathrm{F2}$.
Initially, it was similar to B$_\mathrm{F1}$ and, later, it became
a bright ejection, Ej, which is also visible in the SXI.mpg movie.
In later images, the ejected structure seemed to untwist and turn
South. The similar observed structure of B$_\mathrm{F1}$ and
B$_\mathrm{F2}$ and their rapid appearance, followed by slower
motions, indicate that both could correspond to brightenings of
portions of the filament rather than to mass motions between 1 and
2. Therefore, we will compute the velocity and acceleration of Ej
starting from 2 (marked in Figure~\ref{F-eruption2_images} by a
triangle).

For the measurements we extracted a slice from each image
delimited by the white lines in Figure~\ref{F-eruption2_images}
and computed spatial profiles as total of the pixel values over
the width of each slice. A set of the spatial profiles is shown in
Figure~\ref{F-eruption2_fit}a as a gray scale image in which
motions of each feature are represented by a bright line, whose
instant slope is the velocity. The leftmost horizontal sections
correspond to the initial static condition. The measurements refer
to the foremost edge of the ejection at its earliest appearance
(the triangles in Figure~\ref{F-eruption2_images}).

The onset of the acceleration phase was poorly resolved. The
ejection accelerated for a short time around 07:43, from $v_0 = 0$
to $v_1 \approx 450$~km~s$^{-1}$. Subsequent apparent deceleration
was mainly due to the turn of the ejection towards the South. No
accelerating features, related to E2, were seen afterwards. Our
fit (see Section~\ref{S-kinematics_method}) is shown with white
dashed lines in Figure~\ref{F-eruption2_fit}a and with the white
arrow in Figure~\ref{F-eruption2_images}, where it acceptably
matches the ``nose'' of the ejecta. Acceleration and velocity are
plotted in Figure~\ref{F-eruption2_fit}b,\,c. The acceleration
phase was co-temporal with the microwave burst and exceeded
2~km~s$^{-2}$; deviations of the real shape from the assumed
smooth Gaussian would imply a higher value. The apparent ``jump''
of the ejection (line 1\,--\,2 in Figure~\ref{F-eruption2_fit}a)
from the initial zero speed to $\approx 300$~km~s$^{-1}$ (in the
plane of the sky) could really produce a shock.

The measured velocity and acceleration match those of E2 in
Figure~\ref{F-homol_eruptions}e and f up to its latest
manifestation at 07:49, when it moved along closed loops toward
their West bases. This direction was different from the
orientation of CME1. The properties of E2 show its confined
character. This is incompatible with the suggestion of
\inlinecite{KumarManoUddin2011} about the role of this eruption in
the formation of the 20 November magnetic cloud.

\subsection{Eruption of the Main Filament at 07:56 (E3)}

Comparison of the edge of the main filament in
Figure~\ref{F-eruption2_images} with the reference cross shows that
its slow motion started at about 07:42, being probably triggered by
eruption E2. Then, the filament accelerated around 07:56. With
excellent coalignment and temporal coverage of TRACE images,
measurements are hampered by the very low contrast of the filament.
On the other hand, the filament appears irregularly in H$\alpha$
images, which suffer from atmospheric effects. These problems
complicated the measurements.

\begin{figure} % {12}
  \centerline{\includegraphics[width=0.6\textwidth]
   {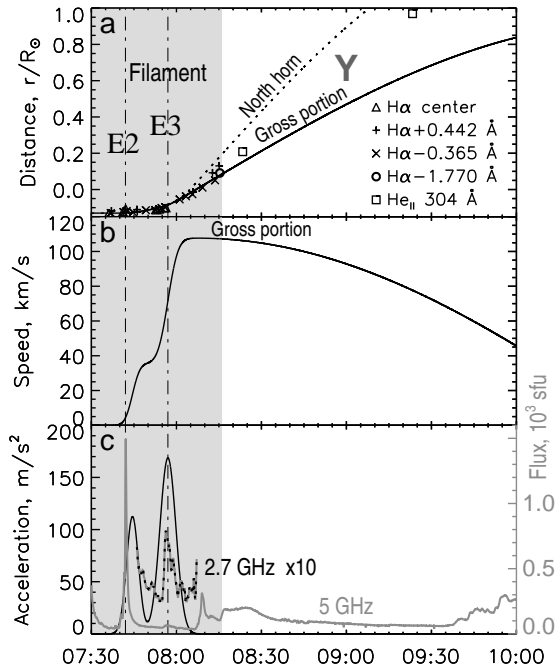}
  }
  \caption{Apparent plane-of-sky kinematic plots of the main
eruption at 07:56. E2 and E3 denote the times of the events
according to Figure~\ref{F-goes_rstn}d. Apparent deceleration
after 08:20 is due to projection shrinkage of the dark Y-feature
moving along a curved trajectory (a circular trajectory is assumed
for simplicity) and is not shown on the acceleration plot.}
  \label{F-eruption3_plots}
  \end{figure}

Using the fact that the filament accelerated twice at about 07:42
and 07:56, we adjusted the parameters of the two acceleration
episodes to follow the darkest filament part in both TRACE and
H$\alpha$ images. We tuned the direction of the measurements from
$-31^{\circ}$ to $-37^{\circ}$ relative to the West. The result is
shown in Figures \ref{F-ha_images} and \ref{F-eruption3_images}
with the white arc, whose shape corresponds to the initial
filament. The kinematic plots are presented in
Figure~\ref{F-eruption3_plots} with solid lines (extensions after
08:16 will be discussed later). The symbols show straightforward
measurements.

When the filament started to move, its lamination became more
pronounced. Some threadlike features lagged behind the gross part,
while some others started to accelerate later, but outran the
gross part. This picture, along with partial jet-like eruptions,
does not correspond to what is expected if the filament was pulled
up as a passive structure inside an expanding larger magnetic flux
rope. Fit in Figure~\ref{F-eruption3_plots} presents the averaged
kinematics of the eruptive filament.

\begin{figure} % {13}
  \centerline{\includegraphics[width=0.6\textwidth]
   {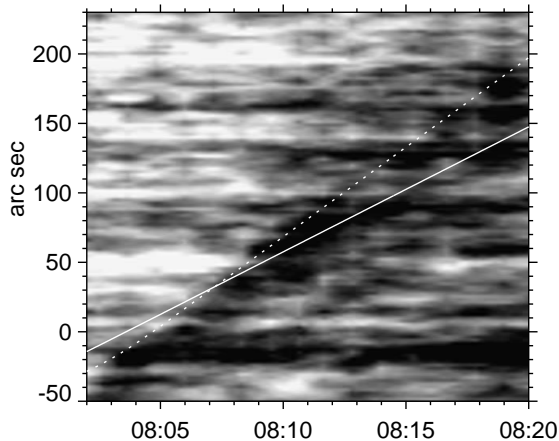}
  }
  \caption{A set of spatial profiles of the main eruption computed
from the red-wing H$\alpha$ images parallel to the North horn of
the Y-like darkening ($-17^{\circ}$ relative to the western
direction, the white line in Figure~\ref{F-eruption3_images}d).
The dotted line fits the fastest motion. The solid line
corresponds to the fit in Figure~\ref{F-eruption3_plots} related
to the direction of the filament's top ($-37^{\circ}$).}
  \label{F-eruption3_red}
  \end{figure}

We also measured the motion of the fastest filament's part
F1$^{\prime \prime}$ (Figures \ref{F-eruption3_images}d and
\ref{F-bifurcation}) detectable in the red-wing H$\alpha$ images.
To catch most parts of the expanding filament, a wide band
inclined by $-17^{\circ}$ to the West was used in the measurements
(Figure~\ref{F-eruption3_images}d). This direction corresponds to
the North horn of the Y-like darkening. A set of the measured
spatial profiles is shown in Figure~\ref{F-eruption3_red}. The
fastest part of the eruptive filament had a nearly constant
velocity of about 210~km~s$^{-1}$ (dotted line) after the
collision with region Rb. The solid line corresponds to the
filament's top measured at $-37^{\circ}$ (the maximum velocity
$\approx 110$~km~s$^{-1}$, Figure~\ref{F-eruption3_plots}b). The
irregular appearance of the ragged F1$^{\prime \prime}$ does not
allow us to plot its kinematic properties in detail, as we did for
the filament's top. The maximum acceleration of $\gsim
500$~m~s$^{-2}$ occurred after 08:02, presumedly when the filament
touched a coronal structure above region Rb.

The kinematic analysis of filament F1 excludes its direct
participation in CME2; although the filament accelerated rather
close to the onset time of CME2 \cite{Gopal2005c}, their speeds
differed by an order of magnitude. However, motion of the filament
is consistent with that of the Y-like darkening inferred from the
SPIRIT He~{\sc ii} 304~\AA\ images (squares in
Figure~\ref{F-eruption3_plots}a). The motion of F1$^{\prime
\prime}$ (from Figure~\ref{F-eruption3_red}) is shown with the
dotted line in this plot.

An apparent shrinkage, due to projection, of the Y-feature in
Figure~\ref{F-spirit}d\,--\,f suggests that its motion is nearly
parallel to the photosphere. In this case, its plane-of-sky
velocity is affected by the cosine of its position angle. Assuming
the surface velocity to be constant, we can extend the plots in
Figure~\ref{F-eruption3_plots} to the South-West up to the limb.
The extrapolated fit of the filament gross portion is shown by the
white arcs in Figure~\ref{F-spirit} that match the middle part of
the Y-darkening. The dotted fit of the fastest filament's part in
Figure~\ref{F-eruption3_plots}a reaches the limb at about 09:07,
this corresponds to the North horn of the Y-feature in
Figure~\ref{F-spirit}e.

\section{Magnetic Fields}
 \label{S-magnetic_fields}

\subsection{Pre-eruptive Filament}
 \label{S-pre-eruptive_filament}

Increasing evidence confirms that filaments are directly related
to coronal magnetic flux-ropes (see, \textit{e.g.},
\opencite{Mackay2010}). Imaging observations reveal prominences
and filaments whose plasmas emit or absorb radiation thus making
them most accessible indicators of magnetic flux-ropes in the
corona. A loss of equilibrium of a flux-rope shows up as a
filament eruption. An eruption appearing before the onset of a
flare or CME is a first manifestation of a solar storm.

The height of a filament above the photosphere is a crucial
parameter to characterize its stability. Models of equilibrium of
flux ropes in the corona imply the existence of a critical value
of the total electric current and its relation with a critical
height of a filament
\cite{vanTendKuperus1978,MolodenskyFilippov1987,FilippovDen2001,DemoulinAulanier2010}.
The critical height, $h_\mathrm{c}$, characterizes the size scale
of the ambient coronal field.

It is not easy to measure the height of a filament observed on the
solar disk. One of few possible ways is to use information about
the tilt of the symmetry plane of a filament with respect to the
vertical by means of the technique proposed by
\inlinecite{d'Azambuja1948}, and then to estimate the height of
the filament from its observed width at any time.
\inlinecite{ZagnetkoFilippovDen2005} found that the symmetry plane
of a filament nearly coincided with the neutral surface of the
coronal potential magnetic field or the surface that passes
through the apex of magnetic field arches. Therefore, a potential
extrapolation can be used to find the tilt and height of a
filament.

\begin{figure}%[h] % {14}
     \centerline{\hspace*{0.015\textwidth}
               \includegraphics[width=0.6\textwidth,clip=,
                 bb=7 5 240 210]
        {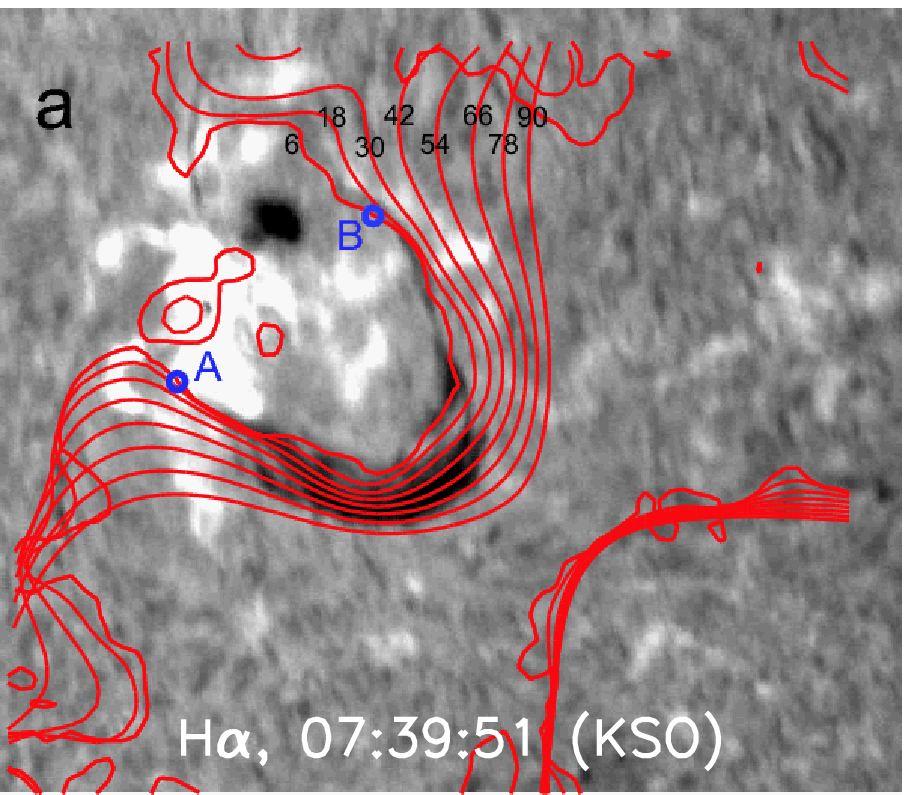}
               \hspace*{-0.01\textwidth}
               \includegraphics[width=0.4\textwidth,clip=]
        {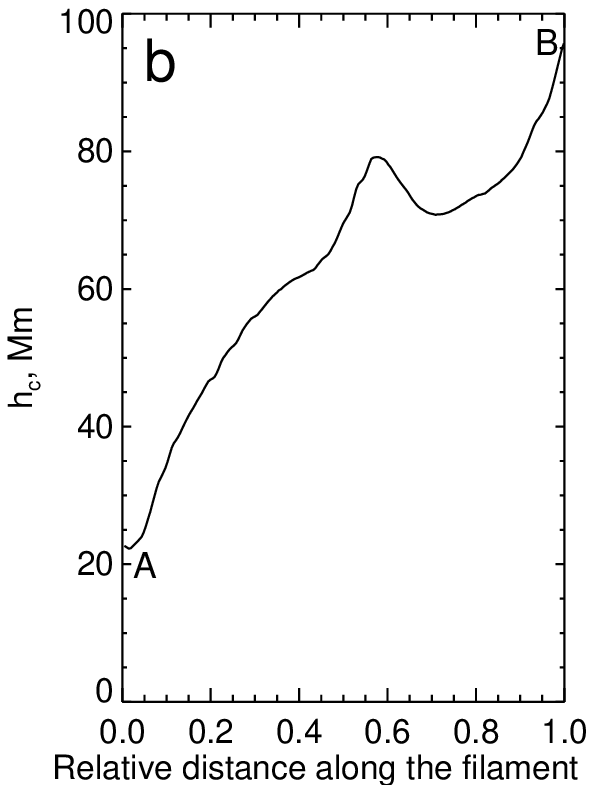}
              }
\caption{(a)~Polarity inversion lines at different heights above
the photosphere in steps of 12 Mm starting from 6 Mm (extrapolated
from the MDI magnetogram observed at 07:59) overlaid on top of the
KSO H$\alpha$ filtergram (07:39:51). (b)~Values of the critical
height $h_{\mathrm{c}}$ along the filament from point A to point B
in panel (a). The horizontal axis shows a dimensionless relative
distance measured from point A (0) to B (1).
        }
 \label{F-filament_nl}
   \end{figure}

We have used the MDI magnetogram at 07:59 as boundary condition
for a potential field extrapolation \cite{FilippovDen2001}. The
KSO H$\alpha$ filtergram observed at 07:39:51 was carefully
coaligned with the magnetogram using sunspots and bringing bright
flocculus kernels into coincidence with magnetic elements of the
network. Figure~\ref{F-filament_nl}a shows the H$\alpha$ image
overlaid with a set of polarity inversion lines calculated for
different heights above the photosphere starting from 6 Mm in
steps of 12~Mm. The inner edge of the filament nearly coincides
with the 6~Mm inversion line, while its outer edge corresponds to
a height of 78~Mm. The set of lines in Figure~\ref{F-filament_nl}a
constitutes a neutral surface of the magnetic field. It is
inclined in the middle part by $\approx 60^{\circ}$ to the
photosphere. Figure~\ref{F-neutral_surface} presents the neutral
surface in more detail.

Figure~\ref{F-filament_nl}b shows values of the critical height
calculated in the way described by \inlinecite{FilippovDen2001}
along the filament from point A to point B in
Figure~\ref{F-filament_nl}a. The critical height was minimum in
the southern part of the filament between small sunspots of
opposite polarities, close to point A. The gradient of the
magnetic field was high there and this site was the best place for
initiation of an eruption. Even though the height of the filament
was low (below the instability threshold,
Figure~\ref{F-filament_nl}), the eruption started in this part. In
the North segment, the critical height was maximum; while the
filament was low. This segment was very stable and did not erupt.
The middle part of the filament had both largest width and height.
It was close to the limit of stability. A small disturbance could
be sufficient to drive an eruption.

\subsection{Variations of Magnetic Fields on the Photosphere}
 \label{S-magnetic_field_variations}

\begin{figure} % {15}
  \centerline{\includegraphics[width=\textwidth]
   {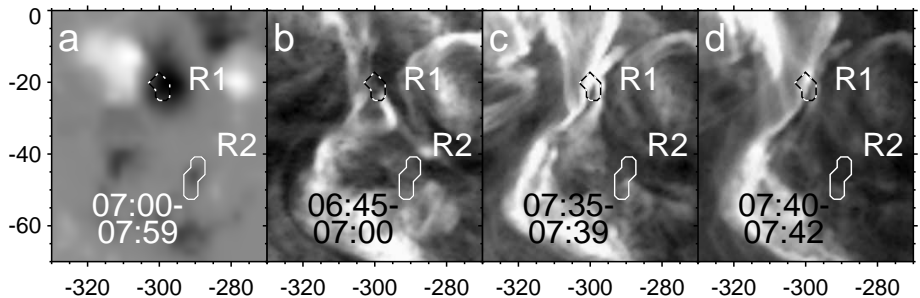}
  }
  \caption{Contours of regions R1 and R2
($3\sigma$ level) overlaid on a magnetogram averaged along 1~hour
(a) and averaged TRACE images in 171~\AA\ obtained before episode
E1 (b), after E1 (c), and in 195~\AA\ averaged during episode E2
(d).}
  \label{F-variable_regions}
  \end{figure}

\begin{figure} % {16}
  \centerline{\includegraphics[width=0.7\textwidth]
   {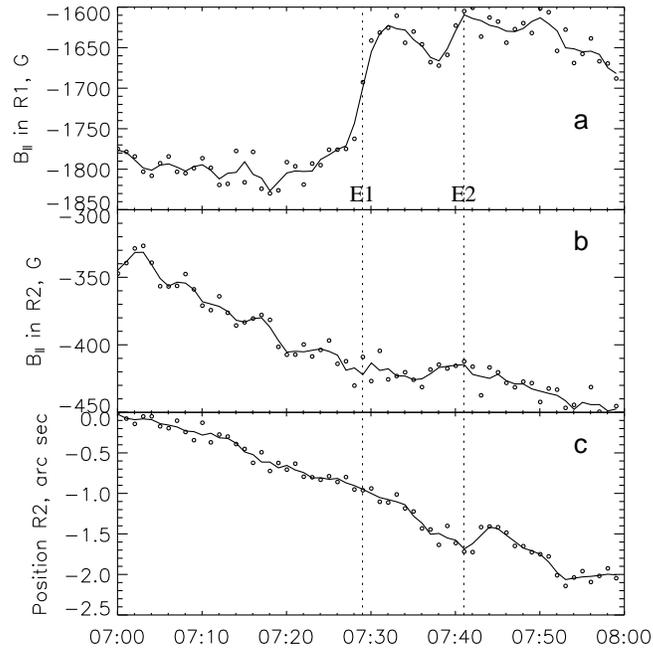}
  }
  \caption{Variations of the line-of-sight magnetic component
$B_{\parallel}$ in regions R1 (a) and R2 (b, see
Figure~\ref{F-variable_regions}) and the displacement of the
magnetic element R2 (c).}
  \label{F-mdi_variations}
  \end{figure}

\inlinecite{Srivastava2009} and \inlinecite{Chandra2010} reported
significant variations of the magnetic fields in AR~501 during a
few days preceding the event. To complete this picture, we
analyzed a set of 60 1-min line-of-sight SOHO/MDI magnetograms
between 07:00 to 08:00, using a variance technique
(\inlinecite{Grechnev2003}; MDI magnetograms were analyzed with
this technique by \inlinecite{Kundu2001} and
\inlinecite{Meshalkina2009}). The solar rotation was compensated
in all the magnetograms, and then the standard deviation was
computed along each pixel in the data cube. Regions exceeding the
$3\sigma$ level in the resulting two-dimensional map present most
variable areas in the magnetograms. The largest changes in the
line-of-sight component ($B_{\parallel}$) correspond to regions R1
and R2 near the filament (Figure~\ref{F-variable_regions}).
Indeed, the TRACE images in Figures~\ref{F-variable_regions}b,\,c
show drastic changes in the corona above R1 during episode E1.
Variations in R2 might have affected a filament barb, but changes
near R2 in the TRACE images in
Figures~\ref{F-variable_regions}b--d are inconclusive.
Quantitative parameters of R1 and R2 are presented in
Figure~\ref{F-mdi_variations}.

The $B_{\parallel}$ component in region R1 changed significantly
around 07:27 and 07:38, just before episodes E1 and E2. Region R2
was a magnetic element moving South by $\approx 2^{\prime \prime}$
from 07:00 to 08:00. A deviation of its motion occurred at about
07:41, at the onset of the eruptive event E2. The $B_{\parallel}$
component in region R2 gradually increased, probably indicating
variations of the slope of magnetic field lines. Regions R1 and R2
were close to point A in Figure~\ref{F-filament_nl}a,
\textit{i.e.}, the South-East end of the filament. As shown in
Section~\ref{S-pre-eruptive_filament}, this part of the filament
was nearly unstable. Thus, the photospheric motions could trigger
eruptions E1 and E2. In turn, disturbances produced by eruption E2
could drive eruption E3 of the major central part of the filament
that was close to the limit of stability.

\subsection{Neutral Surface}

\begin{figure} % {17}
  \centerline{\includegraphics[width=0.58\textwidth,clip=,
                 bb=40 34 306 312]
   {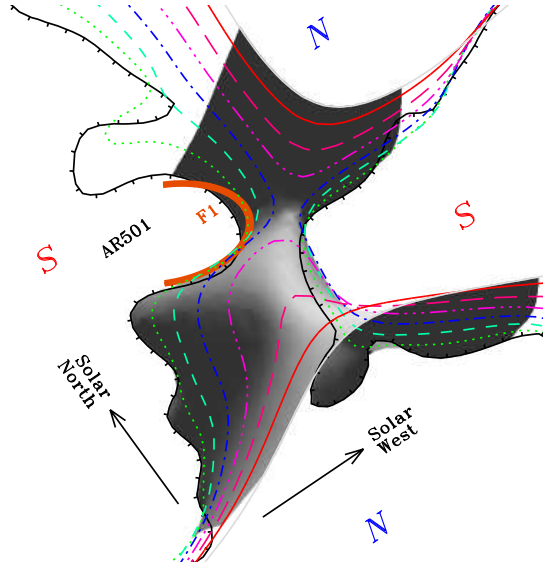}
  }
  \caption{Magnetic neutral surface in the corona computed in the potential
approximation. The orange thick arc denotes the pre-eruptive
filament F1. The projection corresponds to a synoptic magnetogram
rotated by $75^{\circ}$ around the X axis (East--West) and by
$35^{\circ}$ around the Z axis (perpendicular to the photospheric
X--Y plane). Neutral lines of the same color and style correspond
to equal heights in steps of 36 Mm. The blue `N' (positive) and
red `S' (negative) denote predominant magnetic polarities on the
photosphere. Minor magnetic islands are not shown. The
photospheric basis of the shaded neutral surface is from $L =
-18.7^{\circ}$ to $29.1^{\circ}$ (Carrington longitude) and from
$\varphi = -40.9^{\circ}$ to $8.2^{\circ}$ (latitude),
\textit{i.e.}, $\approx 580 \times 600$ Mm$^2$.}
  \label{F-neutral_surface}
  \end{figure}

Threads of filaments are known to be located along inversion
lines, \textit{i.e.}, loci of null radial magnetic component, $B_r
= 0$. By combining a set of magnetic neutral lines calculated at
different heights, one gets a neutral surface. The shape of the
neutral surface is determined by large-scale connections in a
magnetic complex consisting of ARs 501, 503, and their environment
(Figure~\ref{F-overview}a). The neutral surface computed in the
potential approximation \cite{Rudenko2001} from the MDI
magnetogram observed at 07:59 is shown in
Figure~\ref{F-neutral_surface}. The neutral lines at eight height
levels are shown with different colors and line styles. The
surface had a saddle shape resembling a hyperbolic paraboloid. The
\url{nl_movie1.gif} movie facilitates understanding the complex
arrangement of the neutral lines at different heights in the
corona (heliocentric distances are shown in the lower right
corner). A topological discontinuity in $B_r$ occurred at
$1.2015R_\odot$. The projected location of the saddle structure
was close to the bifurcation region Rb.

Having a moderate velocity, an eruptive filament lifts off
presumably along a neutral surface (\citeauthor{Filippov2001},
\citeyear{{Filippov2001},{Filippov2002}};
\opencite{Filippov2008}). This ensures its integrity and
determines the trajectory, which is sometimes erratic. When an
eruption acquires significant kinetic energy, its motion becomes
independent of coronal structures. The shape of the neutral
surface was a topological obstacle for the moving filament. It had
to undergo dramatic changes after passing through the saddle part
of the configuration shown in Figure~\ref{F-neutral_surface}. A
significant part of the magnetic field in the filament could
reconnect with external magnetic fields. This process could result
in magnetic field lines passing inside the eruptive filament and
ending far on the solar surface. Part of the filament plasma moved
along these field lines.

The saddle shape of the neutral surface indicates the presence of
a topological singularity in the large-scale magnetic
configuration. Here the singularity is a null point initially
located at a height of $\sim\,$100~Mm, as discussed in paper~III.
Region Rb was close to the point where the spine field line
leaving the null point entered the photosphere. The projected
positions of the null point, the discontinuity of the $B_r$
component, and region Rb were close to each other.

\section{Absorption in H$\alpha$ and EUV and Mass Estimations}
 \label{S-masses}

The fact that the filament was dark in the H$\alpha$ line
indicates a temperature $< 2 \times 10^4$~K. Emission from such a
filament is a sum of the chromospheric radiation, partly
attenuated by the filament, and its own emission due to
collisional or radiative excitation of hydrogen atoms. A
stationary filament is usually modeled as a vertical
plane-parallel slab with its height being larger than its width.
When a filament activates and loses its shape, it is reasonable to
consider it as a cloud. Several authors \cite[and
others]{Mein1996,MolownyHoras1999,Heinzel2003a} developed non-LTE
models to relate the brightness in the H$\alpha$ line of such a
cloud against the background solar disk to its parameters such as
the density of neutral atoms, temperature, pressure, and
velocities of turbulent and macroscopic motions.

\subsection{Mass of the Pre-eruptive Filament}

The pre-eruptive filament in Figure~\ref{F-ha_images} (row I) had
a projected distance between ends of about 100 Mm and a width of
up to 37 Mm. We estimated the filament mass (without its static
North-West part) at this stage from its opacity by using an
approximate method based on large grids of non-LTE
magneto-hydrostatic models of cool filament-like structures
\cite{Heinzel2003b}. The average brightness ratio of the filament
to its surroundings at 07:30 was $0.85 \pm 0.05$. To find the
optical thickness in the H$\alpha$ line center, we firstly
calculated the parametric dependence of the total absorption in
the H$\alpha$ line within the filter bandwidth on the optical
thickness in the line center. The parameters were the bulk
velocity of the filament, $V$, its turbulent velocity,
$V_\mathrm{turb}$, and temperature $T$. Based on these data and
assuming $T \approx 6000-10000$~K, $V=0$, $V_\mathrm{turb} <
10$~km~s$^{-1}$, and taking into account the filter bandwidth, we
have estimated the optical thickness of the filament in the
H$\alpha$ line center to be $\tau_0 = 0.45 \pm 0.1$.

The dimensions of the filament and its thickness along the line of
sight are necessary for further considerations. The heights of the
filament edges were estimated by comparing the visible filament
with the height distribution of the magnetic neutral lines (see
Figure~\ref{F-filament_nl}a). Assuming the edges of the filament
to coincide with the neutral lines at the same height, we found a
height of 6~Mm for the lower edge, 66\,--\,76 Mm for the upper
edge, and adopted an average height of 70 Mm. The brightness
distribution across the broadest part of the filament can be
reproduced if the filament is considered as a slab with a width of
12 Mm, height of 66 Mm, and an angle to the line of sight of
$\approx 23^\circ$. With these parameters, the average
line-of-sight thickness is $\approx 30$ Mm.

Another assumption concerns the filament's temperature, which
determines its ionization state and the Doppler width of the
H$\alpha$ line; the latter can be increased by turbulent motions.
With a temperature of the filament of $T = 8000$~K (a middle value
for the model used) and zero turbulent velocity, the mass of the
initial filament\footnote{Hereafter the error ranges correspond to
uncertainties in opacity measurements. Masses estimated in terms
of this approximation agree with those estimated by means of
different methods within a factor of 2. The accuracy of relative
variations of masses estimated from single-wavelength images is
considered to be considerably higher.} is $M_{\mathrm{H}\alpha} =
(2-4) \times 10^{15}$~g. This estimate refers to the cold filament
body. It is embraced by a hotter plasma shell, which is a
transition region between the filament and the ambient hot corona.
The shell (sometimes termed the `EUV filament channel') is not
visible in H$\alpha$ but is considered to contribute up to
50\,--\,100\%  of the mass of an erupting filament
\cite{AulanierSchmieder2002}. Taking this contribution into
account, we estimate the total initial mass of the static filament
to be $M_{\mathrm{fil}} = (4-6) \times 10^{15}$~g.

The filament was also visible as a dark absorbing feature in the
TRACE 171~\AA\ and 195~\AA\ images (Figure~\ref{F-ha_images}, left
column), and the opacity in EUV can be used to estimate its mass
as well \cite{AnzerHeinzel2005}. At the assumed temperature of
8000~K, the darkening of a filament in coronal EUV lines can be
due to two factors: \textit{i}) absorption of EUV continuum
radiation from underlying coronal layers by neutral hydrogen and
helium, and \textit{ii}) the absence of the coronal emission from
the volume of the filament (`volume blocking' effect). The latter
fraction depends on the heights of the lower and upper edges of
the filament estimated above. Assuming that the corona is uniform
and the scale height is 40 Mm, the filament mass estimated from
EUV opacity is $(1-2) \times 10^{15} $~g, somewhat less than the
estimate from the H$\alpha$ opacity. This is probably due to
bright loops located above the filament and nearby, whose presence
causes overestimation of background and underestimation of the
optical thickness of the filament.

\subsection{Y-like Darkening at 304~\AA}

The filament accelerated around 07:56, and after 08:01 some of its
parts appeared in the red channel. Between 08:07 and 08:10 (E4A)
the filament reached region Rb (Figure~\ref{F-bifurcation}) where
it bifurcated and presumedly separated into a multitude of pieces
detectable in both the blue and red wings, and after 08:16 it
completely disappeared from the H$\alpha$ line.

The Y-like darkening visible from 08:23 to 09:54 in the SPIRIT
304~\AA\ images (Figures~\ref{F-spirit}d\,--\,f) had no detectable
counterparts in the H$\alpha$ channels. The area of the darkening
at 304~\AA\ was maximum at 09:07 (Figure~\ref{F-spirit}e), when
the average ratio of its brightness to the background was 0.84.
The horizontal velocity of the major part of the cloud at that
time was between 110 and 60~km~s$^{-1}$
(Figure~\ref{F-eruption3_plots}a,b).

Analysis of absorption mechanisms in the He~{\sc ii} 304~\AA\ line
\cite{Grechnev2008} shows that a darkening in this line can be due
to two processes: \textit{i})~non-resonant absorption of incident
continuum emission in hydrogen and helium, and
\textit{ii})~resonant scattering of the He~{\sc ii} emission
irradiated by the whole solar hemisphere on the He~{\sc ii} ions
in the cloud. The latter mechanism alone provides a depression of
$2\pi/4\pi = 0.5$. With a line-of-sight velocity of the cloud
$>50$~km~s$^{-1}$, the Doppler shift of the He~{\sc ii} 304~\AA\
line in the cloud from the emission line diminishes the resonant
scattering. The absence of the cloud in any of the H$\alpha$
channels indicates its rather high vertical velocity. Thus, the
darkening in the He~{\sc ii} 304~\AA\ line was mainly due to the
non-resonant absorption.

The average depressed brightness within the darkening of 0.84
corresponds to the optical thickness of $\tau_{304} = 0.17$.
Assuming the non-resonant absorption mechanism, with an average
absorption cross section of $\sigma_{304} = 5.5 \times
10^{-19}$~cm$^2$ for the $90\%\ \mathrm{H} +10\%\ \mathrm{He}$
mixture \cite{AnzerHeinzel2005}, we obtain the total column number
of atoms $N = 3\times 10^{17}$~cm$^{-2}$. With a thickness of the
absorbing layer of the order of 10~Mm, this corresponds to a
reasonable gas density of $n = 3.6 \times 10^8$~cm$^{-3}$.
Integration over the total area of the absorbing feature gives a
cloud mass of $M_{304} = (2-4) \times 10^{15}$~g from the 304~\AA\
data.

Though no darkening was visible in the SPIRIT 175~\AA\ coronal
images, their processing revealed absorption with an average
optical thickness of $\tau_{175} = 0.02-0.2$. The mass of the
cloud estimated from the 175~\AA\ data is $M_{175} = (1-9) \times
10^{15}$~g with a middle value of $5 \times 10^{15}$~g, which
agrees with the $M_{304}$ within the accuracy of the estimation.
These estimates refer to the total mass of the cloud including the
surrounding EUV filament channel, because EUV radiation is
absorbed with nearly the same cross section by plasmas with
temperatures of up to $8 \times 10^5$~K.

In summary, the average mass of the Y-like cloud absorbing EUV
emission, $(2-4) \times 10^{15}$~g, was close to the mass of the
pre-eruptive filament visible in the H$\alpha$ line. Minor
fractions of the initial mass could be carried away by the ejecta
which erupted at 07:41 and by fastest filament fragments (possible
absorbing material above the limb or behind it was not visible in
the image used in the measurements). Heated material invisible in
EUV can contribute to the total mass. Thus, cool material of the
Y-like cloud inherited the major part of material of the main
filament.

\section{Summary and Discussion}
\label{S-discussion}

Variations of the photospheric magnetic fields suggest that
eruption E1 at 07:29 as well as eruption E2 at 07:41 might have
been triggered by photospheric motions. The E1 event was
presumedly associated with a partial eruption from the easternmost
region of filament F1 and the launch of CME1 in the southeastern
direction. The eruption manifested in short HXR and microwave
bursts and an SXR enhancement with an M1.2 level (which was not
reported).

Episode E2 started with rapid brightenings of long segments in the
South-South-East part of the filament visible in TRACE and SXI
images (similar brightenings of eruptive filaments are often
observed). The average propagation speed of the brightening was
$\geq 550$~km~s$^{-1}$. These circumstances suggest heating in the
brightened part of the filament from an initial temperature of $<
2 \times 10^4$~K up to $\gsim 1$~MK. Heating could be caused by
some process (possibly associated with magnetic reconnection) in
the filament or below it rapidly ignited over a long distance by
an MHD disturbance. A volumetric fraction of the heated part is
unknown, skin-heating is probable \cite{Grechnev2006}. Heating
must result in a rapid increase of the plasma pressure, $2nkT$, by
about two orders of magnitude. Thus, heating alone could cause the
plasma ejection along the magnetic field.

The impulsive eruption E2 started by 07:41:27, before the peak of
the related microwave burst (Figure~\ref{F-goes_rstn}d).
Acceleration of the ejecta reached $\approx 2$~km~s$^{-2}$ at
about 07:43 to reach its final speed of $\approx 450$~km~s$^{-1}$
in the plane of the sky. The microwave burst was also short and
co-temporal with the acceleration pulse. The SXR emission reached
the M3.2 level. The succession of the phenomena suggests that the
flare processes were induced by eruption E2. Observations do not
support a suggestion of \inlinecite{KumarManoUddin2011} that `the
energy release at the first stage of the M3.2 flare
(\textit{i.e.}, E1) triggered the first ejection (E2)'. As
Figure~\ref{F-goes_rstn}d shows, the major energy release of event
E1 that was observed in HXR and microwave bursts ceased well
before the onset of eruption E2. As mentioned, this eruption was
probably triggered by photospheric motions.

Eruption E2 disturbed filament F1, which probably was in a
metastable state. Its apparent twitch in the H$\alpha$ images that
started just after E2 suggests perturbations of plasma flows in
the filament. The filament was initially inclined by $\sim
23^{\circ}$ to the line of sight and started to lift off following
the neutral surface. It moved up and South-West and underwent, at
least, two acceleration phases. The first acceleration pulse
peaked at 07:44:30 ($\approx 110$~m~s$^{-2}$), slightly later than
the acceleration of E2, up to a speed of about $35$~km~s$^{-1}$.
The second acceleration pulse peaked at about 07:57 (episode E3;
$\approx 170$~m~s$^{-2}$) and resulted in a final speed of
$\approx 110$~km~s$^{-1}$. These plane-of-sky velocities agree
with the line-of-sight ones implied by the H$\alpha$ wings. The
latest observation of filament fragments in the far-blue H$\alpha$
wing ($\lambda = 6562.8 -1.770$~\AA) at 08:15:20 (Figures
\ref{F-overview}b and \ref{F-eruption3_images}d) suggesting
upwards velocities of 64\,--\,96~km~s$^{-1}$ also agrees with the
overall kinematics. A weak microwave response to the significant
acceleration during episode E3 was certainly present
(Figure~\ref{F-goes_rstn}d). The close temporal correspondence
between the acceleration episodes of the filament and the flare
bursts provides further support for the active roles of the
filament itself and plasma flows in the filament.

The temporal correspondence between the acceleration of a CME and
flare emissions has been found by \inlinecite{Zhang2001} and
\citeauthor{Temmer2008} (\citeyear{Temmer2008};
\citeyear{Temmer2010}). Reconnection in a sheared arcade
significantly increases the poloidal magnetic field in the
developing flux rope that was quantitatively confirmed by
\inlinecite{Qiu2007}. Due to the curvature of the flux rope, the
toroidal force propelling its expansion increases. After impulsive
acceleration, the poloidal `mainspring' relaxes and the toroidal
force ceases, while the flux rope continues to expand with an
increased speed. This booster mechanism (see, \textit{e.g.},
\opencite{Vrsnak2008}) seems to be commonly present in
flare-related eruptions. This probably occurred in the faint
episode E3, when the filament impulsively accelerated.

The accelerated filament laminated. Its parts showed rather wide
acceleration temporal ranges, maximum accelerations, and speeds.
The strongest acceleration of $\gsim 500$~m~s$^{-2}$ and highest
speed ($\approx 210$~km~s$^{-1}$) were measured along the North
horn of the Y-darkening, where acceleration peaked after 08:02.

The direction of the filament's motion crossed an unavoidable
topological obstacle. As mentioned, a rising eruptive filament is
known to successively reproduce in its motion the contours of the
same neutral line of $Br$ changing with height. The topology of
the neutral lines broke in the saddle region of the neutral
surface (Figure~\ref{F-neutral_surface}), and the filament had no
chance to retain its former configuration and integrity after
passing through this region. Having encountered this obstacle
above region Rb at about 08:07 (episode E4A), the filament
apparently bifurcated into two components F1$^{\prime}$ and
F1$^{\prime \prime}$. Both upwards and downwards motions of these
components are indicated by the red and blue H$\alpha$ wings. The
collision probably resulted in reconnection between magnetic
structures of the filament with environment magnetic fields.

This assumption is supported by the correspondence between the
H$\alpha$ and SXR flare-like response in the bifurcation region
and the HXR peaks E4A and E4B in active region 501 in the interval
from 08:07 to 08:13 because the pre-eruptive filament does not
seem to be connected to this region. After that the bifurcated
filament apparently transformed into two wide diverging jets
constituting the Y-like cloud carrying predominantly cool filament
material to remote sites on the solar surface. The SPIRIT images
show them at 304~\AA\ up to the South-West of the limb
(08:23\,--\,09:55). The cloud consisted of dispersed fragments of
the filament with wide ranges of speeds and directions of motion
within a sector of about $60^{\circ}$ in the plane of the sky and
some vertical scatter. The transformation of the filament into the
Y-like cloud is confirmed by the correspondence between their
masses, as well as their common kinematics. Neither off-limb
extensions of the Y-like cloud, nor a corresponding slow CME are
detectable. Plasma flows from this cloud away from the Sun are not
excluded, while probably most of the mass eventually fell back on
the solar surface far from the eruption site.

In the outlined scenario, reconnection increases the poloidal flux
of an ascending flux rope and completes its formation pushing the
rope into the interplanetary space. In the 18 November event, a
portion of the poloidal flux possibly escaped with the Y-like
cloud coming from the filament. Most of its mass was not ejected
into the interplanetary space. The poloidal flux of this structure
could partly be lost in magnetic reconnection with coronal
magnetic fields and partly carried away as an almost empty
structure invisible in white light. This circumstance might
account for the significant excess of the reconnected magnetic
flux with respect to the MC found by \inlinecite{Moestl2008} and
attributed by the authors to reconnection between the MC and
interplanetary magnetic fields. Moreover, the unusual scenario of
the filament eruption revealed by our analysis makes relating the
flare reconnection flux with the 20 November MC questionable.

The probable scenario of sequential eruptions and subsequent
transformation of the filament in the 18 November 2003 event looks
unusual, but this event was not exceptional in this respect. A
similar anomalous eruption was discussed by
\inlinecite{Grechnev2008}. The transit of an eruptive filament
through a coronal magnetic null point (see, \textit{e.g.},
\opencite{GaryMoore2004}) can transform the filament into an
eruptive jet \cite{{Meshalkina2009}, {Filippov2009}}.
\inlinecite{Grechnev2011_AE} presented other candidates for
anomalous eruptions and showed their probable correlation with the
so-called negative radio bursts (decreases of the total microwave
flux below a quasi-stationary level) without a one-to-one
correspondence. A negative burst could not be observed in the 18
November event, because the anomalous eruption was followed by
ongoing flaring. The authors suggested anomalous eruptions to be
favored by complex magnetic configurations and surrounding of the
active region by others, and did not expect an anomalous eruption
of a quiescent filament beyond an activity complex. The clearest
example of anomalous eruption was observed by SDO/AIA on 7 June
2011 between 06:15 and 10:00 when dispersed absorbing fragments of
a disintegrating filament were well visible even in the 193~\AA\
images without subtraction. The ``destruction'' of the magnetic
structure of an eruptive filament in such anomalous eruptions
deserves the attention of future studies.

The anomalous character of the main eruption in the 18 November
2003 event brings the attention to CME2
(Figure~\ref{F-overview}c). Its faint, fast outer halo envelope
crossing a distorted streamer could be a trace of a shock wave
\cite[section 4.3.2]{Grechnev2011_I} excited by one of sharp
eruptions. The initiation time of the inner component estimated by
\inlinecite{Gopal2005c} suggests its development in the complex E4
event. The appearance of the inner component corresponds to the
pointing of the pre-eruptive filament F1 and its initial
non-radial motion. Remarkable is a radial structure of the inner
component suggesting its development from an arcade which was
initially located above the filament or its trajectory, and then
expanded being forced to erupt. Neither cavity nor core were
clearly evident, whereas a threadlike core (former filament) is
typically a brightest CME component. These features of CME2 are
consistent with our conclusion that the bulk of the filament
material had not left the Sun (\textit{cf.}
\citeauthor{Grechnev2008}, \citeyear{Grechnev2008};
\citeyear{Grechnev2011_AE}). We will consider CME2 in more detail
in paper~II.

A possible extra eruption seems to have occurred between 08:07 and
08:17 close to the disk center. This eruption is suggested by the
central dimming in the images of GOES/SXI
(Figure~\ref{F-ha_sxi}i), SPIRIT 175~\AA\
(Figure~\ref{F-spirit}b\,--\,c), and EIT
(Figure~\ref{F-overview}c) as well as the `disconnection' of the
bifurcation region from the flare site shown by the H$\alpha$ and
HXR time profiles between events E4B and E4C in
Figure~\ref{F-ha_sxi}g. Similar manifestations of disconnection
were considered by \inlinecite{Kundu2001} and
\inlinecite{GrechnevWhiteKundu2003} as indications of changes in
the magnetic connectivity during solar flares. The presumable
eruption between 08:07 and 08:17 could result in one extra CME,
which might not be detected in LASCO images. This issue will be
addressed in papers~II and III.

We have found out what happened to the main filament mass, but it
is not so clear what happened to its magnetic flux. Part of it
might be carried away frozen in the Y-like inheritor of the
bifurcated filament. Some part of the magnetic flux probably
reconnected during the interaction with the bifurcation region.
The final outcome is uncertain. Anyway, these circumstances make
doubtful a simple scenario, in which the magnetic cloud hitting
Earth is considered as a stretched flux rope formed from a
structure initially associated with the pre-eruption filament F1.
\inlinecite{Grechnev2005} concluded that in the 18 November event
`the eruptive filament probably failed to become the CME core',
but this result was not considered by other authors who analyzed
the causes of the extreme geoeffective disturbance and its
possible solar source. They will be revisited in paper~IV.

The challenges revealed in preceding studies of the 20 November
magnetic cloud and its presumable solar source, extended by our
results, are as follows:

\begin{enumerate}

 \item
Very strong magnetic field in the MC that was close to a record
value.

 \item
Different handedness of the MC and the presumed solar source.

 \item
Different orientations of the magnetic field in the MC and the
presumed solar source ($> 90^{\circ}$, \opencite{Moestl2008}).

 \item
Implication of CME1 looks doubtful, because a similar CME in the
stronger 17 November event was not geoeffective.

 \item
Responsibility of the moderately fast coreless CME2 for the
severest geomagnetic storm seems to be unlikely.

\end{enumerate}

Items (2) and (3) indicate that the commonly assumed association
of the MC with the U-shaped filament in AR~501 is questionable.
The suggestions of the eruption which presumedly occurred between
08:07 and 08:17 imply the development of an extra CME. If this
hypothetical CME expanded exactly earthward from the solar disk
center within a very narrow cone, then the Thomson-scattered light
was meager to be detected by LASCO. Such atypically weak expansion
of the CME must result in very strong magnetic field inside due to
magnetic flux conservation. Paper~II addresses further suggestions
of this possible CME.

\section{Conclusion}
\label{S-conclusion}

To complement the picture of the complex 18 November 2003 event
established in preceding papers, we have studied a succession of
eruptions and the corresponding flare episodes E1, E2, E3, and E4
in the interval 07:29 to 08:30. We have measured the kinematic
properties of eruptive filaments and found that they underwent
impulsive acceleration episodes, which were temporally close to
microwave/HXR bursts. The eruptions rapidly expanded and lost
opacity which made difficult their subsequent observation.
Nevertheless, it was possible to reveal their further evolution.
Our results and their implications are as follows.

\begin{enumerate}

\item
 \textit{Triggers and mechanisms of eruptions.}
Probable primary triggers of the partial eruptions at 07:29 (E1)
and 07:41 (E2) were photospheric motions. Eruption E2 presumably
occurred due to rapid heating of a long filament segment probably
caused by an MHD disturbance. The corresponding sharp pressure
increase in the filament could induce an ejection of some part of
its material along the magnetic field. In turn, the disturbance
produced by eruption E2 most likely destabilized filament F1 and
triggered its eruption at 07:56 (E3).

\item
 \textit{Filament eruptions and CME initiation.}
The eruptions presumably initiated the CMEs. Development of the
eruptions after their triggers by the factors listed above appears
to be self-sufficient so that outer drivers do not seem to be
required. This circumstance is important for models of eruptions
and initiation of CMEs, while eruptive filaments are often
considered to be passive structures being pulled upwards by
larger-scale magnetic ropes.

\item
 \textit{Anomalous eruption.}
The major eruption was anomalous; after the initial lift-off, the
eruptive filament bifurcated and had not left the Sun as a whole,
while its main mass probably fell back on the solar surface. This
anomaly occurred due to the collision around 08:07 of the eruptive
filament with an unavoidable topological discontinuity---a
magnetic null point. Being not able to pass through the obstacle
keeping integrity, the filament bifurcated. Flare episodes E4A
(08:09) and E4B (08:12) responded to this process.

\item
 \textit{The problem of the geomagnetic impact.}
Relation to the 20 November 2003 magnetic cloud is doubtful for
the 07:29 eruption presumedly associated with CME1 and the 07:41
eruption, which most likely had not produced any CME at all. The
transformation of the eruptive filament, whose main part had not
left the Sun, contradicts simple considerations of the magnetic
cloud hitting the Earth as a stretched magnetic flux-rope formed
from a structure initially associated with the pre-eruption
filament. On the other hand, a possible additional eruption, which
appears to have occurred between 08:07 and 08:17 close to the
solar disk center, might be implicated in the development of a
structure responsible for the geomagnetic superstorm. Further
substantiation of this scenario will be presented in Papers II to
IV.

\end{enumerate}

\begin{acks}

We thank Viktoria Kurt for the CORONAS-F/SONG data, L.~Kasha\-pova
and S.~Kalashnikov for the assistance in data processing, and
I.~Kuzmenko for useful discussions. We are grateful to an
anonymous reviewer for valuable recommendations to improve the
paper. We thank the instrumental teams of the Kanzelh{\"o}he Solar
Observatory; TRACE and CORONAS-F missions; MDI, EIT, and LASCO on
SOHO (ESA \& NASA); the USAF RSTN Radio Solar Telescope Network;
and the GOES satellites for the data used here.

This study was supported by the Russian Foundation of Basic
Research under grants 11-02-00757, 11-02-01079, 12-02-00008,
12-02-92692, and 12-02-00037, the Program of basic research of the
RAS Presidium No.~22, and the Russian Ministry of Education and
Science under State Contract 16.518.11.7065. The research was also
partly supported by the European Commission's Seventh Framework
Programme (FP7/2007-2013) under the grant agreement eHeroes
(project No. 284461), \url{www.eheroes.eu}.

\end{acks}

\end{article}


\begin{thebibliography}{}

\bibitem[\protect\citeauthoryear{Anzer and Heinzel}{2005}]{AnzerHeinzel2005}
Anzer, U., Heinzel, P.: 2005, \apj\ \textbf{622}, 714.

\bibitem[\protect\citeauthoryear{Aulanier and Schmieder}{2002}]{AulanierSchmieder2002}
Aulanier, G. Schmieder, B.: 2002, \aap\ \textbf{386}, 1106.

\bibitem[\protect\citeauthoryear{Bogachev {\it et al.}}{2009}]{Bogachev2009}
Bogachev, S.A., Grechnev, V.V., Kuzin, S.V., Slemzin, V.A.,
Bugaenko, O.I., Chertok, I.M.: 2009, {\it Solar Sys. Res.} {\bf
43}, 143.

\bibitem[\protect\citeauthoryear{Brueckner {\it et al.}}{1995}]{Brueckner1995}
Brueckner, G.E., Howard, R.A., Koomen, M.J., Korendyke, C.M.,
Michels, D.J., Moses, J.D., Socker, D.G., Dere, K.P., \textit{et
al.}: 1995, \solphys\ \textbf{162}, 357.

\bibitem[\protect\citeauthoryear{Carrington}{1859}]{Carrington1859}
Carrington, R.C.: 1859, \mnras\ {\bf 20}, 13.

\bibitem[\protect\citeauthoryear{Chandra {\it et al.}}{2010}]{Chandra2010}
Chandra R., Pariat, E., Schmieder, B., Mandrini, C.H., Uddin, W.:
2010, \solphys\ \textbf{261}, 127.

\bibitem[\protect\citeauthoryear{Chertok and Grechnev}{2005}]{ChertokGrechnev2005}
Chertok, I.M., Grechnev, V.V.: 2005, \textit{Astron. Rep.}
\textbf{49}, 155.

\bibitem[\protect\citeauthoryear{d'Azambuja and d'Azambuja}{1948}]{d'Azambuja1948}
d'Azambuja, M., d'Azambuja, L.: 1948, \textit{Ann. Obs. Paris,
Meudon} \textbf{6}, Fasc. VII.

\bibitem[\protect\citeauthoryear{Delaboudini\`ere {\it et al.}}{1995}]{Delab1995}
Delaboudini\`ere, J.-P., Artzner, G.E., Brunaud, J., Gabriel, A.H.,
Hochedez, J.-F., Millier, F., Song, X.Y., Au, B.,  \textit{et al.}:
1995, \solphys\ \textbf{162}, 291.

\bibitem[\protect\citeauthoryear{D{\'e}moulin and Aulanier}{2010}]{DemoulinAulanier2010}
D{\'e}moulin,~P., Aulanier,~G.: 2010, \apj\ {\bf 718}, 1388.

\bibitem[\protect\citeauthoryear{Echer, Gonzalez, and Tsurutani}{2008}]{Echer2008}
Echer, E., Gonzalez, W.D., and Tsurutani, B.T.: 2008, \grl\ {\bf
350}, 6.

\bibitem[\protect\citeauthoryear{Filippov and Den}{2001}]{FilippovDen2001}
Filippov, B.P., Den, O.G.: 2001, \jgr\ \textbf{106}, 25177.

\bibitem[\protect\citeauthoryear{Filippov, Golub, and Koutchmy}{2009}]{Filippov2009}
Filippov, B., Golub, L., Koutchmy, S.: 2009, \solphys\ {\bf 254},
259.

\bibitem[\protect\citeauthoryear{Filippov, Gopalswamy, and Lozhechkin}{2001}]{Filippov2001}
Filippov, B.P., Gopalswamy, N., Lozhechkin, A.V.: 2001, \solphys\
\textbf{203}, 119.

\bibitem[\protect\citeauthoryear{Filippov, Gopalswamy, and Lozhechkin}{2002}]{Filippov2002}
Filippov, B.P., Gopalswamy, N., Lozhechkin, A.V.: 2002,
\textit{Astron. Rep.} \textbf{46}, 417.

\bibitem[\protect\citeauthoryear{Filippov and Koutchmy}{2002}]{Filippov2008}
Filippov, B., Koutchmy, S.: 2008, \ag\ \textbf{26}, 3025.

\bibitem[\protect\citeauthoryear{Gary and Moore}{2004}]{GaryMoore2004}
Gary, G. A., Moore, R. L.: 2004, \apj\ {\bf 611}, 545.

\bibitem[\protect\citeauthoryear{Gopalswamy {\it et al.}}{2005a}]{Gopal2005a}
Gopalswamy, N., Barbieri, L., Cliver, E.W., Lu, G., Plunkett, S.P.,
Skoug, R.M.: 2005a, \jgr\ \textbf{110}, A09S00.
%doi:10.1029/2005JA011268.

\bibitem[\protect\citeauthoryear{Gopalswamy {\it et al.}}{2005b}]{Gopal2005b}
Gopalswamy, N., Yashiro, S., Liu, Y., Michalek, G., Vourlidas, A.,
Kaiser, M.L., Howard, R.A.: 2005b, \jgr\ \textbf{110}, A09S15.
%doi:10.1029/2004JA010958.

\bibitem[\protect\citeauthoryear{Gopalswamy {\it et al.}}{2005c}]{Gopal2005c}
Gopalswamy, N., Yashiro, S., Michalek, G., Xie, H., Lepping, R.P.,
Howard, R.A.: 2005c, \grl\ \textbf{32}, L12S09.

\bibitem[\protect\citeauthoryear{Grechnev}{2003}]{Grechnev2003}
Grechnev, V.V.: 2003, \solphys\ \textbf{213}, 103.

\bibitem[\protect\citeauthoryear{Grechnev, White, and Kundu}{2003}]{GrechnevWhiteKundu2003}
Grechnev, V.V., White, S.M., Kundu, M.R.: 2003, \apj\ {\bf 588},
1163.

\bibitem[\protect\citeauthoryear{Grechnev {\it et al.}}{2005}]{Grechnev2005}
Grechnev, V.V., Chertok, I.M., Slemzin, V.A., Kuzin, S.V., Ignat'ev,
A.P., Pertsov, A.A., Zhitnik, I.A., Delaboudini\`ere, J.-P.,
Auch\`ere, F.: 2005, \jgr\ \textbf{110}, A09S07.

\bibitem[\protect\citeauthoryear{Grechnev {\it et al.}}{2006}]{Grechnev2006}
Grechnev, V.V., Uralov, A.M., Zandanov, V.G., Baranov, N.Y.,
Shibasaki, K.: 2006, \pasj\ {\bf 58}, 69.

\bibitem[\protect\citeauthoryear{Grechnev {\it et al.}}{2008}]{Grechnev2008}
Grechnev, V.V., Uralov, A.M., Slemzin, V.A., Chertok, I.M.,
Kuzmenko, I.V., Shibasaki, K.: 2008, \solphys\ \textbf{253}, 263.

\bibitem[\protect\citeauthoryear{Grechnev {\it et al.}}{2011a}]{Grechnev2011_AE}
Grechnev, V.V., Kuzmenko, I.V., Chertok, I.M., Uralov, A.M.: 2011a,
\textit{Astron. Rep.} \textbf{55}, 637.

\bibitem[\protect\citeauthoryear{Grechnev {\it et al.}}{2011b}]{Grechnev2011_I}
Grechnev, V.V., Uralov, A.M., Chertok, I.M., Kuzmenko, I.V.,
Afanasyev, A.N., Meshalkina, N.S., Kalashnikov, S.S., Kubo, Y.:
2011b, \solphys\ {\bf 273}, 433.

\bibitem[\protect\citeauthoryear{Handy {\it et al.}}{1999}]{Handy1999}
Handy, B.N., Acton, L.W., Kankelborg, C.C., Wolfson, C.J., Akin,
D.J., Bruner, M.E., Caravalho, R., Catura, R.C., \textit{et al.}:
1999, \solphys\ \textbf{187}, 229.

\bibitem[\protect\citeauthoryear{Heinzel, Anzer, and Schmieder}{2003a}]{Heinzel2003a}
Heinzel, P., Anzer, U., Schmieder, B.: 2003a, \solphys\ {\bf 216},
159.

\bibitem[\protect\citeauthoryear{Heinzel {\it et al.}}{2003b}]{Heinzel2003b}
Heinzel, P., Anzer, U., Schmieder, B., Schwartz, P.: 2003b,
\textit{ESA SP-535}, 447.

\bibitem[\protect\citeauthoryear{Hill {\it et al.}}{2005}]{Hill2005}
Hill, S.M., Pizzo, V.J., Balch, C.C., Biesecker, D.A., Bornmann, P.,
Hildner, E., \textit{et al.}: 2005, \solphys\ \textbf{226}, 255.

\bibitem[\protect\citeauthoryear{Ivanov, Romashets, and Kharshiladze}{2006}]{Ivanov2006}
Ivanov, K.G., Romashets, E.P., Kharshiladze, A.F.: 2006, {\it
Geomag. and Aeron.} {\bf 46}, 275.

\bibitem[\protect\citeauthoryear{Kumar, Manoharan, and Uddin}{2011}]{KumarManoUddin2011}
Kumar, P., Manoharan, P.K., Uddin, W.: 2011, \solphys\ {\bf 271},
149.

\bibitem[\protect\citeauthoryear{Kundu {\it et al.}}{2001}]{Kundu2001}
Kundu, M.R., Grechnev, V.V., Garaimov, V.I., White, S.M.: 2001,
\apj\ {\bf 563}, 389.

\bibitem[\protect\citeauthoryear{Kuznetsov {\it et al.}}{2011}]{Kuznetsov2011}
Kuznetsov, S.N., Kurt, V.G., Yushkov, B.Y., Kudela, K., Galkin,
V.I.: 2011, \solphys\ {\bf 268}, 175.

\bibitem[\protect\citeauthoryear{Lin {\it et al.}}{2002}]{Lin2002}
Lin, R.P., Dennis, B.R., Hurford, G.J., Smith, D.M., Zehnder, A.,
Harvey, P.R., \textit{et al.}: 2002, \solphys\ \textbf{210}, 3.

\bibitem[\protect\citeauthoryear{Lui}{2011}]{Lui2011}
Lui, A.T.Y.: 2011, \ssr\ {\bf 158}, 43.

\bibitem[\protect\citeauthoryear{Mackay {\it et al.}}{2010}]{Mackay2010}
Mackay, D.H., Karpen, J.T., Ballester, J.L., Schmieder, B.,
Aulanier, G.: 2010, \ssr\ {\bf 151}, 333.

\bibitem[\protect\citeauthoryear{Mari{\v c}i{\'c} {\it et al.}}{2004}]{Maricic2004}
Mari{\v c}i{\'c}, D., Vr{\v s}nak, B., Stanger, A.L., Veronig, A.:
2004, \solphys\ {\bf 225}, 337.

\bibitem[\protect\citeauthoryear{Mari{\v c}i{\'c} {\it et al.}}{2007}]{Maricic2007}
Mari{\v c}i{\'c}, D., Vr{\v s}nak, B., Stanger, A.L., Veronig, A.M.,
Temmer, M., Ro{\v s}a, D.: 2007, \solphys\ {\bf 241}, 99.

\bibitem[\protect\citeauthoryear{Mein {\it et al.}}{1996}]{Mein1996}
Mein, N., Mein, P., Heinzel, P., Vial, J.-C., Malherbe, J.M.,
Staiger, J.: 1996, \aap\ \textbf{309}, 275.

\bibitem[\protect\citeauthoryear{Meshalkina {\it et al.}}{2009}]{Meshalkina2009}
Meshalkina, N. S., Uralov, A. M., Grechnev, V. V., Altyntsev, A.
T., Kashapova, L. K.: 2009, \pasj\ {\bf 61}, 791.

\bibitem[\protect\citeauthoryear{Miklenic {\it et al.}}{2007}]{Miklenic2007}
Miklenic, C.H., Veronig, A.M., Vr{\v s}nak, B., Hanslmeier, A.:
2007, \aap\ \textbf{461}, 697.

\bibitem[\protect\citeauthoryear{Miklenic, Veronig, and Vr{\v s}nak}{2009}]{Miklenic2009}
Miklenic, C.H., Veronig, A.M., Vr{\v s}nak, B.: 2009, \aap\
\textbf{499}, 893.

\bibitem[\protect\citeauthoryear{Molodensky and Filippov}{1987}]{MolodenskyFilippov1987}
Molodensky, M.M., Filippov, B.P.: 1987, \sovast\ \textbf{31}, 564.

\bibitem[\protect\citeauthoryear{M{\" o}stl {\it et al.}}{2008}]{Moestl2008}
M{\" o}stl, C., Miklenic, C., Farrugia, C.J., Temmer, M., Veronig,
A., Galvin, A.B., Vr{\v s}nak, B., Biernat, H.K.: 2008, \ag\
\textbf{26}, 3139.

\bibitem[\protect\citeauthoryear{Molowny-Horas {\it et al.}}{1999}]{MolownyHoras1999}
Molowny-Horas, R., Heinzel, P., Mein, P., Mein, N.: 1999, \aap\
\textbf{345}, 618.

\bibitem[\protect\citeauthoryear{Neupert}{1968}]{Neupert1968}
Neupert, W. M.: 1968, \apj\ \textbf{153}, L59.

\bibitem[\protect\citeauthoryear{Oraevsky and Sobelman}{2002}]{OraevskySobelman2002}
Oraevsky, V.N., Sobelman, I.I.: 2002, {\it Astron. Lett.} {\bf
28}, 401.

\bibitem[\protect\citeauthoryear{Oraevsky {\it et al.}}{2003}]{Oraevsky2003}
Oraevsky, V.N., Sobelman, I.I., Zitnik, I.A., Kuznetsov, V.D.,
Stepanov, A.I., Polishuk, G.M., \textit{et al.}: 2003, \adv\ {\bf
32}, 2567.

\bibitem[\protect\citeauthoryear{Pizzo {\it et al.}}{2005}]{Pizzo2005}
Pizzo, V.J., Hill, S.M., Balch, C.C., Biesecker, D.A., Bornmann, P.,
Hildner, E., \textit{et al.}: 2005, \solphys\ \textbf{226}, 283.

\bibitem[\protect\citeauthoryear{Qiu {\it et al.}}{2007}]{Qiu2007}
Qiu, J., Hu, Q., Howard, T.A., Yurchyshyn, V.B.: 2007, \apj\ {\bf
659}, 758.

\bibitem[\protect\citeauthoryear{Rompolt}{1998}]{Rompolt1998}
Rompolt, B.: 1998, {\it IAU Colloq. 167: New Perspectives on Solar
Prominences} {\bf 150}, 330.

\bibitem[\protect\citeauthoryear{Rudenko}{2001}]{Rudenko2001}
Rudenko, G.V.: 2001, \solphys\ {\bf 198}, 5.

\bibitem[\protect\citeauthoryear{Rudenko and Grechnev}{1999}]{RudenkoGrechnev1999}
Rudenko, G. V., Grechnev, V. V.: 1999, {\it Astronomical Data
Analysis Software and Systems VIII}, ASP Conf. Series, {\bf 172},
421.

\bibitem[\protect\citeauthoryear{Scherrer {\it et al.}} {1995}]{Scherrer1995}
Scherrer, P. H., Bogart, R. S., Bush, R. I., Hoeksema, J. T.,
Kosovichev, A. G., Schou, J., \textit{et al.}: 1995, \solphys\
\textbf{162}, 129.

\bibitem[\protect\citeauthoryear{Slemzin {\it et al.}}{2004}]{Slemzin2004}
Slemzin, V., Chertok, I., Grechnev, V., Ignat'ev A., Kuzin S.,
Pertsov A., Zhitnik I., Delaboudini\`ere J.-P.: 2004, In A.V.
Stepanov, E.E. Benevolenskaya, and A.G. Kosovichev, editors,
Multi-Wavelength Investigations of Solar Activity, {\it Proc. IAU
Symp. 223}, 533.

\bibitem[\protect\citeauthoryear{Slemzin {\it et al.}}{2005}]{Slemzin2005}
Slemzin, V.A., Kuzin, S.V., Zhitnik, I.A., Delaboudiniere, J.-P.,
Auchere, F., Zhukov, A.N., van der Linden, R., Bugaenko, O.I.,
Ignat'ev, A.P., Mitrofanov, A.V., Pertsov, A.A., Oparin, S.N.,
Stepanov, A.I., Afanas'ev, A.N.: 2005, {\it Solar Sys. Res.} {\bf
39}, 489.

\bibitem[\protect\citeauthoryear{Srivastava {\it et al.}}{2009}]{Srivastava2009}
Srivastava, N., Mathew, S.K., Louis, R.E., Wiegelmann, T.: 2009,
\jgr\ \textbf{114}, A03107.
%, doi.org/10.1029/2008JA013845.

\bibitem[\protect\citeauthoryear{Temmer {\it et al.}}{2008}]{Temmer2008}
Temmer, M., Veronig, A.M., Vr{\v s}nak, B., Ryb{\' a}k, J., G{\"
o}m{\" o}ry, J., Stoiser, S., Mari{\v c}i{\' c}, D.: 2008, \apj\
\textbf{673}, L95.

\bibitem[\protect\citeauthoryear{Temmer {\it et al.}}{2010}]{Temmer2010}
Temmer, M., Veronig, A.M., Kontar, E.P., Krucker, S., Vr{\v s}nak,
B.: 2010, \apj\ {\bf 712}, 1410.

\bibitem[\protect\citeauthoryear{van Tend and Kuperus}{1978}]{vanTendKuperus1978}
van Tend, W., Kuperus, M.: 1978, \solphys\ \textbf{59}, 115.

\bibitem[\protect\citeauthoryear{Tsurutani {\it et al.}}{2003}]{Tsurutani2003}
Tsurutani, B.T., Gonzalez, W.D., Lakhina, G.S., Alex, S.: 2003,
\jgr\ {\it A} {\bf 108}, 1268.

\bibitem[\protect\citeauthoryear{Veselovsky {\it et al.}}{2004}]{Veselovsky2004}
Veselovsky, I.S., Panasyuk, M.I., Avdyushin, S.I., Bazilevskaya,
G.A., Belov, A.V., Bogachev, S.A. {\it et al.}: 2004,
\textit{Cosmic Res.} \textbf{42}, 435.

\bibitem[\protect\citeauthoryear{Vr{\v s}nak}{2008}]{Vrsnak2008}
Vr{\v s}nak, B.: 2008, \ag\ {\bf 26}, 3089.

\bibitem[\protect\citeauthoryear{Wang, Zhang, and Shen}{2009}]{WangZhangShen2009}
Wang, Y., Zhang, J., Shen, C.: 2009, \jgr\ {\bf 114}, 10104.

\bibitem[\protect\citeauthoryear{Yashiro {\it et al.}}{2004}]{Yashiro2004}
Yashiro, S., Gopalswamy, N., Michalek, G., St. Cyr, O. C.,
Plunkett, S. P., Rich, N. B., Howard, R. A.: 2004, \jgr\ 109,
A07105.
%, doi:10.1029/2003JA010282

\bibitem[\protect\citeauthoryear{Yermolaev {\it et al.}}{2005}]{Yermolaev2005}
Yermolaev Yu.I., Zelenyi, L.M., Zastenker, G.N., {\it et al.}:
2005, \textit{Geomag. and Aeron.}, \textbf{45}, 20.

\bibitem[\protect\citeauthoryear{Yurchyshyn, Hu, and Abramenko}{2005}]{Yurchyshyn2005}
Yurchyshyn, V., Hu, Q., Abramenko, V.: 2005, \textit{Space
Weather}, \textbf{3}, S08C02.
%, doi:10.1029/2004SW000124.

\bibitem[\protect\citeauthoryear{Zagnetko, Filippov, and Den}{2005}]{ZagnetkoFilippovDen2005}
Zagnetko, A.M., Filippov, B.P., Den, O.G.: 2005, \textit{Astron.
Rep.} \textbf{49}, 425.

\bibitem[\protect\citeauthoryear{Zhang {\it et al.}}{2001}]{Zhang2001}
Zhang, J., Dere, K.P., Howard, R.A., Kundu, M.R., White, S.M.: 2001,
\apj\ {\bf 559}, 452.

\bibitem[\protect\citeauthoryear{Zhitnik {\it et al.}}{2002}]{Zhitnik2002}
Zhitnik, I.A., Bougaenko, O.I., Delaboudini\`ere, J.-P., Ignatiev,
A.P., Korneev, V.V., Krutov, V.V., Kuzin, S.V., Lisin, D.V.,
\textit{et al.}: 2002, {\it Proc. 10th European Solar Physics
Meeting, Prague (ESA SP-506)}, 915.


\end{thebibliography}
\end{document}